\documentclass[3p,12pt]{elsarticle}

\usepackage{graphics}
 \usepackage{graphicx}

\usepackage{amssymb}

\usepackage{lineno}
\usepackage{amsmath}
\usepackage{amssymb}
\usepackage{enumerate}
\usepackage{wrapfig}
\usepackage{caption}
\usepackage{subcaption}
\usepackage{lscape}
\usepackage{float}
\usepackage{fancyhdr}
\usepackage{sidecap}

\biboptions{authoryear}

\journal{European Journal of Mechanics - B/Fluids}

\begin{document}

\begin{frontmatter}



\title{The formation of a bubble from a submerged orifice}
\author[brum]{Jonathan A. Simmons}
\ead{simmonsj@maths.bham.ac.uk}
\author[war]{James E. Sprittles}
\ead{j.e.sprittles@warwick.ac.uk}
\author[brum]{Yulii D. Shikhmurzaev}
\ead{y.d.shikhmurzaev@bham.ac.uk}

\address[brum]{School of Mathematics, University of Birmingham, Edgbaston, Birmingham, B15 2TT, United Kingdom.}
\address[war]{Mathematics Institute, University of Warwick, Coventry, CV4 7AL, United Kingdom.}

\begin{abstract}
The formation of a single bubble from an orifice in a solid surface, submerged in an incompressible, viscous Newtonian liquid, is simulated. The finite element method is used to capture the multiscale physics associated with the problem and to track the evolution of the free surface explicitly. The results are compared to a recent experimental analysis and then used to obtain the global characteristics of the process, the formation time and volume of the bubble, for a range of orifice radii; Ohnesorge numbers, which combine the material parameters of the liquid; and volumetric gas flow rates. These benchmark calculations, for the parameter space of interest, are then utilised to validate a selection of scaling laws found in the literature for two regimes of bubble formation, the regimes of low and high gas flow rates.


\end{abstract}


\end{frontmatter}


\section{Introduction}\label{sec:intro}
The controlled production of small gas bubbles is of critical importance to many processes found in the chemical, petrochemical, nuclear, metallurgical and biomedical industries. Subsequently, a vast body of research exists on the subject \citep{Kuma70,Clif78,Kulk05,Yang07}. Due to the complex mechanisms involved in this fundamental process, the computational, experimental and theoretical literature has focused on the case of generating a bubble by pumping gas through a single formation site, namely a submerged vertical nozzle \citep{Long91,Oguz93,Wong98,Thor07} or, as is the case in this work, an upward facing orifice in a submerged solid surface \citep{Zhan01,Bada07,Gerl07,Das11}.

Some authors have considered the case of inflating a bubble under a constant gas pressure with the volumetric gas flow rate varying as the bubble inflates \citep{Dav60a,Dav60b,Saty69,Lana72}. Others have considered the case where the volumetric gas flow rate is determined by the difference between some ambient pressure away from the bubble and the gas pressure in a chamber that is connected to the formation site \citep{Khur69,Oguz93}. However, the most popular method, and the method employed here, is to apply a constant volumetric gas flow rate through the formation site and allow the gas pressure to vary in time \citep{Wong98,Jami01,Corc06,Gerl07}.

\subsection{Experimental Observations} 
The majority of experimental studies examine a continuous chain of bubbles \citep{Zhan01,Bada07,Das11}. A bubble grows whilst attached to the formation site as gas is pumped through the site at a constant flow rate. As the volume of the bubble increases, the influence of buoyancy becomes more important. The bubble seeks to minimise its surface area at a given volume and so a `neck' develops in the bubble as the longitudinal curvature of the free surface changes sign at some point just above the three phase solid--liquid--gas contact line. The difference in pressure between the base and the apex of the bubble then drives the thinning of this neck as the free surface begins to pinch--off (see Figure~\ref{fig:geometry}). This leads to the eventual break up of the bubble into two parts; a new bubble is released and rises away from the formation site under buoyancy, whilst the residual bubble, that is left attached to the formation site, begins to grow. This process repeats itself, thus producing a chain of bubbles.

The aim of many of these studies has been to find the global characteristics of the formation process, such as the frequency of formation and the volumes of the bubbles that are formed, for a given set of material parameters (e.g. liquid density and viscosity), design parameters (e.g. orifice radius, wettability of substrate) and regime parameters (e.g. gas flow rate). These studies provide potentially useful material for comparing theoretical results with experiments. However, global characteristics of bubble formation in a chain accumulate several complex phenomena, and it is desirable to have a more detailed picture of each of them and of their interaction.

Theoretical and early computational studies focused on the behaviour of a single bubble which grows from some initial state up until the break up of the free surface is approached \citep{Long91,Oguz93,Wong98,Xiao05}. The mathematical difficulties of handling the topological change associated with the complete break up of the bubble prohibited any further progress and so, in this case, the formation time period and the volume of the bubble above the point of minimum neck radius are of interest and have been investigated. Here, the break up is assumed to be a local effect that does not affect the global dynamics of the formation process.

Experimental and theoretical studies identified three regimes for the formation of bubbles under a constant gas flow rate \citep{Mcca71,Zhan01,Bada07}. For a given set of material and design parameters, the first of these regimes occurs at small gas flow rates and is known as the `static' regime. In this regime, the volume of the bubbles formed is independent of flow rate, and therefore a decrease in the flow rate results in an increase in the formation time. Consequently, it is not possible to produce a bubble with a volume smaller than this limiting volume. \citet{Fritz35} suggested that this limiting final volume is proportional to the orifice radius. This, as well as other scaling laws are considered in section~\ref{sec:results}. 

At greater flow rates, bubble formation enters the `dynamic' regime. It is now the formation time that approaches a limiting value, resulting in an increase of bubble volume $V_d$ with flow rate $Q$. In this regime, some authors used spherical bubble models to propose scaling laws for the bubble volume. \citet{Dav60b} and \citet{Oguz93} proposed that $V_d \propto Q^{\frac{6}{5}}$ for bubble formation in an inviscid liquid, whilst, \citet{Wong98} suggested that $V_d\propto Q^{\frac{3}{4}}$ in the case of a highly viscous liquid. 

Finally, under even greater flow rates, bubble formation enters the `turbulent' regime. Here the motion is chaotic with successive bubbles coalescing with each other above the formation site. The transition between these three regimes is dependent on all of the material and design properties involved. For instance, in the case of an orifice, where the liquid phase only partially wets the solid surface and so the contact line is free to move along the solid surface, the wettability of the substrate is important to the formation process \citep{Lin94,Byak03,Gnyl03,Corc06}.  

\subsection{Theoretical Progress}
Under sufficiently small gas flow rates, the initial growth of a single bubble may be accurately modelled as a quasi--static process, an approach that has also been utilised for the description of the formation of liquid drops \citep{Ford48,Thor05}. By assuming that the fluid velocities are negligible, one can derive the Young--Laplace equation, which balances the difference in the gas and hydrostatic pressures with the capillary pressure. This can be solved to find the free surface profile of a bubble at a particular volume. By stipulating a small increase in the bubble volume, a series of successive profiles can be found which are in very good agreement with experiments that describe the initial evolution of a bubble \citep{Marm73,Long91,Gerl05,Lee09,Vafa10a,Vafa10b,Vafa11,Lee12,Lesa13}.

However, once the neck forms and the pinch--off process begins, the liquid velocities adjacent to the point of minimum neck radius are no longer negligible. Due to the dynamics associated with the relatively large liquid velocities involved at this stage, the quasi--static approach is no longer valid and thus, in general, one cannot expect to accurately predict the final volume of a newly formed bubble this way.

In an attempt to describe some aspects of the dynamic problem, early mathematical models of single bubble formation under a constant gas flow rate were based upon global force balances. The first `one--stage' models for highly viscous \citep{Dav60a} and inviscid liquids \citep{Dav60b}, which involved the bubble growing spherically, were developed into various `two--stage' models for highly viscous \citep{Rama69,Gadd86} and inviscid liquids \citep{Wrai71,Buye96} by adding a detachment stage to the spherical expansion stage. Once the spherical bubble had reached a certain volume, it translates away from the formation site but remains in contact with it via a cylindrical column of gas. The bubble is said to detach when the column reaches a certain height. In order to incorporate intermediate values of viscosity, some authors solved a modified Rayleigh--Plesset equation, which describes a spherical bubble oscillating under pressure, by assuming that each element on the free surface is the corresponding sphere with the same longitudinal radius of curvature. From this, the mean gas pressure could be found that was then used to solve an equation for the upwards translation of the bubble \citep{Pinc81,Tera90,Tera93}. Although these semi--empirical models are seen to give good agreement in certain regimes \citep{Kulk05}, to accurately describe the whole parameter space of interest, the use of complex computational techniques is needed.

\subsection{Computational Approaches}
To solve the unsteady free--boundary problem of bubble formation from an orifice or nozzle, subject to the forces of gravity, inertia, viscosity and capillarity, the development of computational fluid dynamics techniques is required. Much of the early work in this direction was concerned with the axisymmetric generation of a single bubble in either the inviscid \citep{Marm76,Hoop86,Tan86,Oguz93,Xiao05,Hig06} or highly viscous flow regime \citep{Wong98,Hig05}, where the boundary integral method can be used to reduce the problem's dimensionality to one. 

Since then, the majority of work on bubble generation has focused on the influence of the material, design and regime parameters on the global characteristics of the flow using the volume--of--fluid method \citep{Ma12}, and its improved variants which utilise level--set methods \citep{Buwa07,Gerl07,Chak09,Ohta11,Chak11,Alba13}. Due to the simple manner in which topological changes are `automatically' handled, such techniques have proved successful at describing many features of the bubble formation phenomenon, including the wake effect of a preceding bubble on subsequent bubbles in a chain, where the size of forming bubbles are assumed sufficiently large and hence the details of how the topological transition (i.e. the break up of the bubble) takes place are relatively unimportant. However, as noted in \citet{Egg08}, there is no guarantee that the physics associated with this transition has been properly accounted for. This aspect becomes more important when the size of forming bubbles is small.
 
Despite the advantages of volume--of--fluid based methods, as a relatively simple way of handling the global dynamics, it is well known that this class of numerical techniques are not well suited to resolving the multiscale physics which becomes critical in `singular' flows, i.e. those in which liquid bodies coalesce \citep{Hopp90,Egg99} or divide \citep{Rayl92,Egg97}. For instance, for a millimetre--sized bubble, experiments are able to resolve the minimum neck radius down to tens of microns \citep{Burt05,Thor07,Bola08,Bola09}, whilst numerical methods have thus far failed to capture these scales and often artificially truncate the simulation far above the scales which are still well within the realm of continuum mechanics. These problems have been highlighted in the work of \citet{Hys09}. 

These deficiencies can be addressed by using the finite element method which has previously been used to capture inherently multiscale flows in dynamic wetting \citep{Wils06,Spri13} and in the coalescence of liquid drops \citep{Spri12b}. Moreover, it can easily be applied to the bubble formation phenomenon so that (a) the parameter--space of interest to technologically--relevant processes can be investigated, safe in the knowledge that all scales of the problem have been accurately resolved, and global characteristics of the flow, such as the formation time and bubble volume, can be extracted and (b) the actual pinch--off of the bubble can be studied and compared to various theories proposed in the literature for `singular' flows \citep{Burt05,Gord05,Gord06,Thor07,Bola08,Quan08,Gord08,Bola09,Gek09,Font11}. Historically, aspects (a) and (b) have been considered by separate scientific communities, but in this work, techniques are used which are able to resolve both features. Here, the focus will be on aspect (a), i.e. a parametric study, with a forthcoming publication studying aspect (b) in detail. 

The outline of the paper is as follows. In section~\ref{sec:probform} the problem formulation is given and three dimensionless parameters are identified. Section~\ref{sec:parameter} then contains the range of physically--realistic values of these parameters that are to be used to investigated each parameter's influence on the bubble formation problem. A brief description of the finite element computational framework used is given in section~\ref{sec:framework}. The results are presented in section~\ref{sec:results} where a comparison with an experiment can be found together with the benchmark calculations, which describe the influence of the dimensionless parameters on the global characteristics of the bubble formation process for the parameter space of interest. These calculations are then used to validate various scaling laws found in the literature. Finally, some concluding remarks can be found in section~\ref{sec:disc}. 

\section{Problem Formulation}\label{sec:probform}
Consider a smooth, horizontal, stationary, impermeable solid surface submerged in a quiescent, incompressible, viscous Newtonian liquid of constant density $\rho$ and dynamic viscosity $\mu.$ The solid surface has a circular orifice of dimensionless radius $r_c$ through which an inviscid gas is pumped at a constant dimensionless volumetric flow rate $Q$ to form a single bubble. The characteristic velocities in the gas and the size of the resulting bubble are assumed to be sufficiently small so that any spatial non--uniformity of the gas pressure in the bubble can be neglected.

The following scales are defined for length $L=\sqrt{\sigma/\rho g},$ velocity $U=\sigma/\mu,$ pressure $\sigma/L=\sqrt{\rho \sigma g},$ time $L/U=\mu/\sqrt{\rho \sigma g}$ and flow rate $L^2 U=\sigma^2/\mu\rho g,$ where $\sigma$ is the surface tension of the gas--liquid interface and $g$ is the acceleration due to gravity. In other words, the scales are based on a regime in which viscous, capillarity and buoyancy forces are of similar magnitude, so that the Bond number ($Bo=\rho g L^2/\sigma$) and capillary number ($Ca=\mu U/\sigma$) are unity. From here on, all quantities are dimensionless unless stated and where dimensional quantities do appear, they are denoted with bars.

The incompressible flow of the liquid is governed by the dimensionless Navier--Stokes equations,
\begin{subequations}\label{eq:NS}
\begin{equation}\label{eq:N1}{\bf{\nabla}}\cdot{\bf{u}}=0,\end{equation}
\begin{equation}\label{eq:N2} \frac{\partial {\bf{u}}}{\partial t} +\left({\bf{u}}\cdot{\bf{\nabla}}\right){\bf{u}}=Oh^2\left({\bf{\nabla}}\cdot{\bf{P}}-{\bf{e}}_z\right), \end{equation}
\end{subequations}
where ${\bf{u}}={\bf{u}}({\bf{r}},t)$ and $p=p({\bf{r}},t)$ are the liquid velocity and pressure, ${\bf{r}}$ is the spatial coordinate vector, $t$ is time, ${\bf{P}}=-p{\bf{I}}+{\bf{\nabla u}}+\left({\bf{\nabla u}}\right)^T$ is the stress tensor, ${\bf{I}}$ is the metric tensor and ${\bf{e}}_z$ is a unit vector in the opposite direction to gravity. The Ohnesorge number is $Oh=\mu/\sqrt{\rho \sigma L}$ which, by substituting the length scale, is, $$Oh=\frac{\mu g^{1/4}}{\rho^{1/4}\sigma^{3/4}}.$$ 

\begin{figure}
\begin{center}
\includegraphics[width=0.5\textwidth]{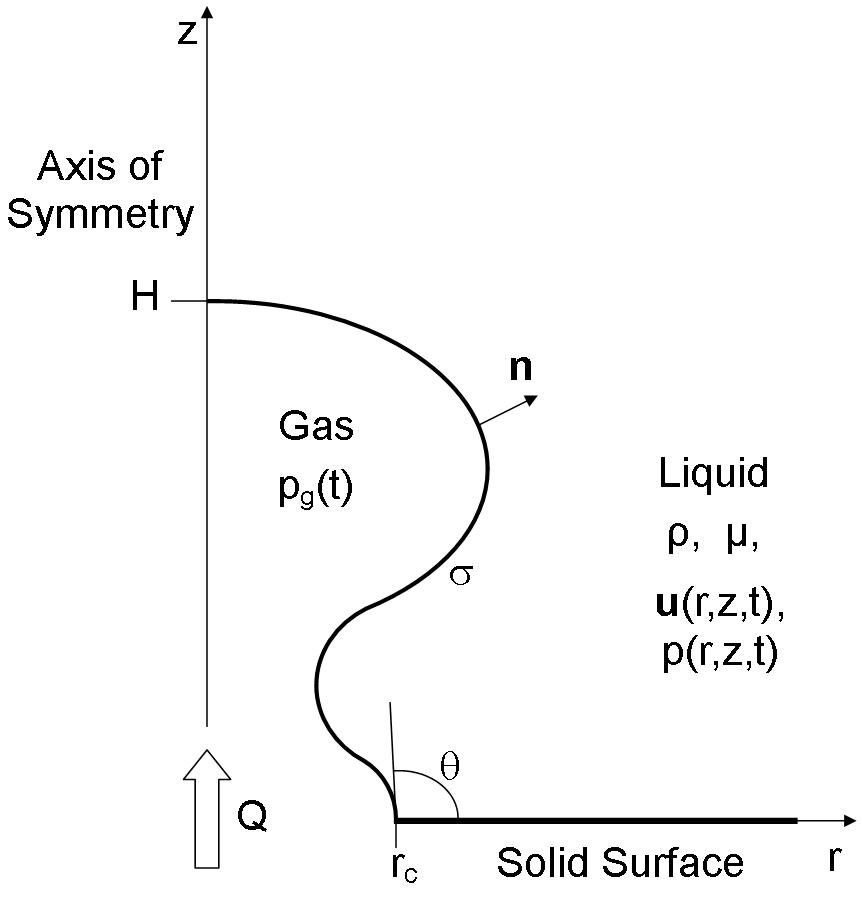}
\caption{A sketch of the flow domain in the $(r,z)$--plane.}
\label{fig:geometry}
\end{center}
\end{figure}

The kinematic boundary condition on the free surface, that states the fluid particles that form the free surface remain there for all time, is,
\begin{equation}\label{eq:kine} \frac{\partial f}{\partial t} +{\bf{u}}\cdot{\bf{\nabla}}f=0,\end{equation}
where $f({\bf{r}},t)=0$ defines the a--priori unknown free surface. The balance of forces acting on an element of the free surface from the liquid and gas phases and neighbouring surface elements manifests itself through two dynamic boundary conditions, the normal and tangential stress conditions,
\begin{equation}\label{eq:normstress} p_g + {\bf{n}}\cdot{\bf{P}}\cdot{\bf{n}}={\bf{\nabla}}\cdot{\bf{n}}, \end{equation}
\begin{equation}\label{eq:tangstress} {\bf{n}}\cdot{\bf{P}}\cdot\left({\bf{I}}-{\bf{n}} {\bf{n}}\right)={\bf{0}},\end{equation}
respectively, where $p_g=p_g(t)$ is the spatially uniform gas pressure, which is determined as part of the solution, and ${\bf{n}}$ is the unit normal to the free surface that points into the liquid phase (see Figure~\ref{fig:geometry}).
The combined no--slip and impermeability condition is applied on the solid surface,
\begin{equation}\label{eq:noslip} {\bf{u}}={\bf{0}}.\end{equation} 

The fluid flow is assumed to be axisymmetric about the vertical axis and therefore the problem is considered in the $(r,z)$--plane where $r$ and $z$ are the respective radial and vertical coordinates  of a cylindrical coordinate system (see Figure~\ref{fig:geometry}). The velocity and pressure fields of the liquid can therefore be rewritten as ${\bf{u}}(r,z,t)=u(r,z,t){\bf{e}}_r+w(r,z,t){\bf{e}}_z$ and $p(r,z,t),$ respectively, where ${\bf{e}}_r$ is the unit coordinate vector in the $r$ direction. The a--priori unknown free surface is now defined as $f(r,z,t)=0.$

The conditions of axial symmetry and smoothness of the velocity field at the axis of symmetry are given by,
\begin{equation} u=0,\qquad \frac{\partial w}{\partial r}=0, \end{equation}
whilst in the far field, there is,
\begin{equation}\label{eq:farfieldbc} u,\,w\rightarrow 0,\qquad r^2 +z^2\rightarrow \infty. \end{equation} 

The normal stress boundary condition (\ref{eq:normstress}) on the free surface is itself a differential equation determining the free surface shape and as such requires its own boundary conditions. Firstly, it is assumed that the liquid phase perfectly wets the solid surface and so the three phase solid--liquid--gas contact line remains pinned to the rim of the orifice for all time, 
\begin{equation} f(r_c,0,t)=0, \end{equation}
where the contact angle $\theta=\theta(t)$ at the pinned contact line is then determined as a part of the solution. Secondly, the free surface is smooth at the bubble apex ($r=0,$ $z=H$) and so,
\begin{equation} {\bf{\nabla}}f\cdot{\bf{e}}_r=0, \end{equation}
is applied there, where $H=H(t)$ is the height of the bubble to also be determined as part of the solution.

For the initial--value problem, the initial velocity field and the initial shape of the bubble must be prescribed. It is assumed that the liquid is initially at rest,
\begin{equation}\label{eq:initu} u(r,z,0)=0,\qquad w(r,z,0)=0,\end{equation}
and that the bubble is initially a spherical cap with volume $V_i.$ The initial height of the bubble $H_i=H(0)$ is the real solution of $\pi H_i\left(3 r_c^2+H_i^2 \right)=6 V_i$ and the radius of the sphere is defined by $R_i=\left(H_i^2+r_c^2\right)/2 H_i.$ The initial free surface shape is then given by,
\begin{equation}\label{eq:fsshape} r^2 +\left(z+R_i-H_i\right)^2 = R_i^2,\qquad 0 \leq z \leq H_i. \end{equation}
In other words, by defining the radius of the orifice and the initial volume of the bubble, the initial shape is fully specified. 

To conclude the problem formulation, an equation governing the volume of the bubble is required in order to employ a constant volumetric gas flow rate. The volume of the bubble is given by,
\begin{equation} V(t)=V_i+Q\,t,\end{equation}
and so if the free surface is described by the function $r=h(z,t)$ where $f(r,z,t)=h(z,t)-r$ then,
\begin{equation}\label{eq:volume} V(t)=\pi\int_{z=0}^{z=H}\,h^2(z,t)\,dz.\end{equation}
The topological change associated with the break up of the bubble, as the simply connected gas phase becomes multiply connected, has not been accounted for in this problem formulation and therefore any simulations will stop short of the break up of the bubble. Each simulation runs until the minimum neck radius $r_{min}=r_{tol},$ where the tolerance $r_{tol}$ is discussed in section~\ref{sec:framework}. 

\section{Parameter Regime}\label{sec:parameter}
The problem can be characterised by the three dimensionless parameters identified in the problem formulation above; the orifice radius $r_c,$ the Ohnesorge number $Oh$ and the volumetric gas flow rate $Q.$ 

To estimate the parameter regime of interest, consider three typical Newtonian liquids; water ($\mu=1$~mPa~s, $\rho=998.2$~kg~m$^{-3}$, $\sigma=73$~mN~m$^{-1}$), a silicone oil ($\mu=0.01$~Pa~s, $\rho=800$~kg~m$^{-3}$, $\sigma=20$~mN~m$^{-1}$) and glycerol ($\mu=1.4$~Pa~s, $\rho=1200$~kg~m$^{-3}$, $\sigma=60$~mN~m$^{-1}$). The respective length scales and Ohnesorge numbers for these three liquids are then $L=2.73$~mm, $1.60$~mm and $2.26$~mm and $Oh=2.24\times 10^{-3},$ $6.26\times 10^{-2}$ and $3.49.$ 

In order to investigate the full effects of these parameters, the system shall be examined using the orifice radii $r_c=0.1$ and $1,$ since it is the generation of bubbles from orifice radii of the order of millimeters and below that are of interest; the Ohnesorge numbers $Oh=2.24\times 10^{-3},$ so that the case of water can be examined explicitly, and also $Oh=10^{-2},$ $10^{-1},$ $1$ and $10,$ to cover the full range of liquid viscosity; and flow rates $10^{-6} \leq Q \leq 0.5$ for $r_c=0.1$ and $10^{-5} \leq Q \leq 15$ for $r_c=1.$ Each simulation will be characterised by the notation $\left(r_c,Oh,Q\right).$  

The initial volume of the bubble $V_i$ is chosen to be sufficiently small enough so that, even though a smaller initial volume would slightly increase the formation time $t_d,$ it would not alter the volume of the newly formed bubble $V_d.$ It was seen for all simulations that an initial contact angle of $\theta(0)=3\pi/4$ is suitable to fulfill this criterion and so the initial volumes for the cases of $r_c=0.1$ and $1$ are $V_i=6.89\times10^{-4}$ and $6.89\times10^{-1},$ respectively.

The length and velocity scales used here are different from those which the reader may have encountered in similar works. A detailed guide on how to rescale the results presented in this manuscript can be found in \ref{app:rescaling}.  
 
\section{A Computational Framework for Bubble Formation}\label{sec:framework}
Due to the mathematical complexity of the bubble formation problem, where gravitational, viscous, inertial and capillarity forces are all present in an unsteady free--boundary problem, a computational method is required to solve the dimensionless system of equations (\ref{eq:NS})--(\ref{eq:volume}). The finite element computational framework used here is based on a formulation described in detail in \citet{Spri12a,Spri13}, where a user--friendly step--by--step guide to its implementation can be found alongside numerous benchmark calculations. This method has already been used to simulate both drop impact phenomena \citep{Spri12c} and, more recently, two--phase coalescence processes \citep{Spri14}. Consequently, the method is only briefly described here and the reader is referred to the aforementioned references for further details.
\begin{figure}[t!]
\begin{center}
	\begin{subfigure}[]{0.5\textwidth}
	\centering
	\includegraphics[width=8cm,keepaspectratio]{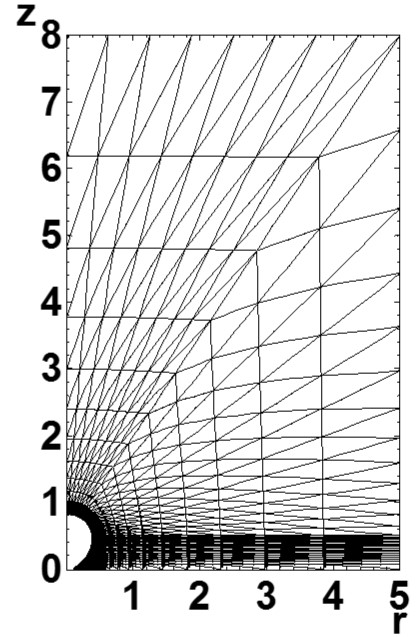}
	\caption{The entire flow domain corresponding to the computed profile in Figure~\ref{fig:newexp2}.}
	\label{fig:newmesha}
	\end{subfigure}
	\qquad
	\begin{subfigure}[]{0.4\textwidth}
	\centering
	\includegraphics[width=5cm,keepaspectratio]{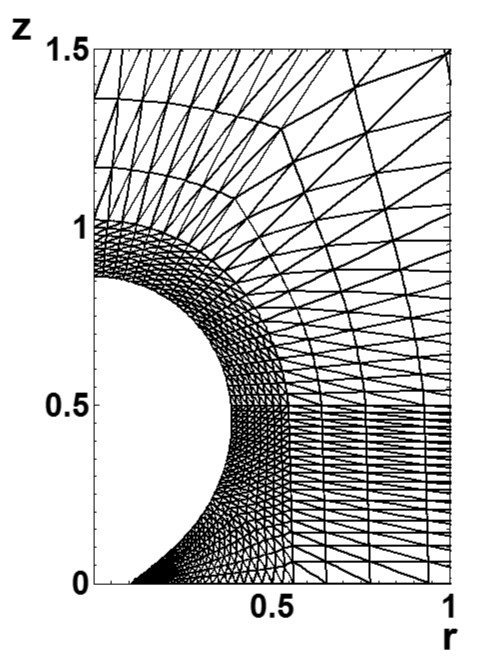}
	\caption{A close up of the computational mesh around the bubble from Figure~\ref{fig:newmesha}.}
	\label{fig:newmeshb}
	\includegraphics[width=5cm,keepaspectratio]{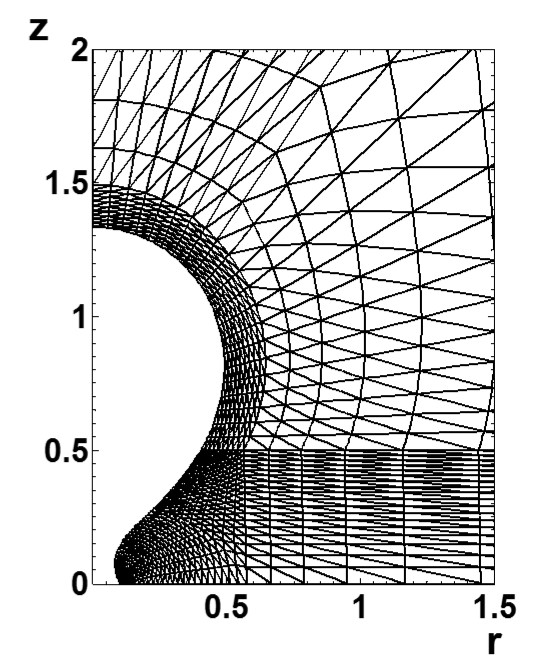}
	\caption{A close up of the computational mesh adjacent to the bubble corresponding to the computed profile in Figure~\ref{fig:newexp4}.}
	\label{fig:newmeshc}
	\end{subfigure}
	\caption{The evolution of the computational mesh for the simulation shown in Figure~\ref{fig:bubble}.}
	\label{fig:mesh}
\end{center}
\end{figure}

The problem is formulated in an infinite domain ($r\geq0,$ $z\geq0$) with boundary conditions (\ref{eq:farfieldbc}) prescribed in the `far--field'. In order to apply the finite element method, the flow domain must be truncated to a finite size. Therefore, a side boundary ($r=R,$ $0\leq z \leq Z$) and a top boundary ($0\leq r \leq R,$ $z=Z$) are constructed on which amended `passive' boundary conditions are applied. To ensure these amended conditions are indeed passive, $R$ and $Z$ are set large enough so that any increase in their values does not affect the growth of the bubble. This is achieved when $R$ is at least five times the maximum width of the bubble and $Z$ is at least four times the maximum height of the bubble. 

As is standard in the finite element method \citep{Gres99}, the system of equations is discretised over a set of nodes that are distributed throughout the flow domain, whilst a finite number of non--overlapping triangular elements are tessellated about these nodes to form a computational mesh (see Figure~\ref{fig:mesh}). The V6--P3 elements are used to approximate velocity quadratically and pressure linearly thus satisfying the Ladyzhenskaya--Babu\u{s}ka--Brezzi condition \citep{Babu72}, whilst curved elements allow the free surface to also be captured with a quadratic approximation. 

An arbitrary Lagrangian--Eulerian mesh design based on the `method of spines' \citep{Kist83} is used where nodes are distributed along the free surface to track its evolution explicitly, whilst the nodes in the bulk move in a prescribed manner that ensures the elements remain non--degenerate. 

\begin{figure}[t!]
\begin{center}
	\begin{subfigure}[]{0.4\textwidth}
	\centering
	\includegraphics[width=7.5cm,keepaspectratio]{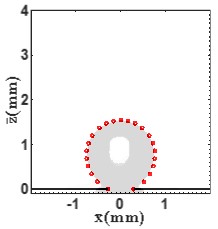}
	\caption{$\bar{t}_e=9.60\times10^{-2}$~s and $\bar{t}=0.683$~s.}
	\label{fig:newexp1}
	\end{subfigure}
	\qquad
	\begin{subfigure}[]{0.4\textwidth}
	\centering
	\includegraphics[width=7.5cm,keepaspectratio]{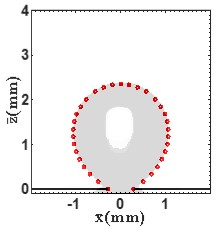}
	\caption{$\bar{t}_e=1.380$~s and $\bar{t}=1.967$~s.}
	\label{fig:newexp2}
	\end{subfigure}
	
	\begin{subfigure}[]{0.4\textwidth}
	\centering
	\includegraphics[width=7.5cm,keepaspectratio]{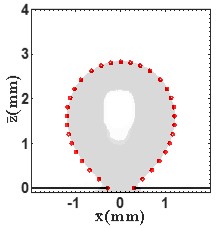}
	\caption{$\bar{t}_e=2.340$~s and $\bar{t}=2.927$~s.}
	\label{fig:newexp3}
	\end{subfigure}
	\qquad
	\begin{subfigure}[]{0.4\textwidth}
	\centering
	\includegraphics[width=7.5cm,keepaspectratio]{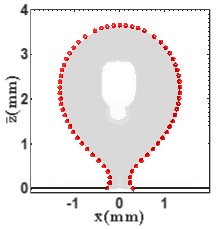}
	\caption{$\bar{t}_e=3.550$~s and $\bar{t}=4.137$~s.}
	\label{fig:newexp4}
	\end{subfigure}
	\caption{The superposition of computational free surface profiles and experimental images for the case $(r_c,Oh,Q)=(0.106,2.24\times10^{-3},5.11\times10^{-6}).$ The experimental images are reproduced from \citet{Bari13}. The dimensional experimental $\bar{t}_e$ and simulation $\bar{t}$ times are given with each subfigure.}
	\label{fig:bubble}
\end{center}
\end{figure}

Figure~\ref{fig:mesh} shows the evolution of the computational mesh for the simulation shown in Figure~\ref{fig:bubble}. Figure~\ref{fig:newmesha} shows that the finite element method is ideally suited to deal with the bubble formation process as the mesh can be refined by placing smaller elements adjacent to the bubble where additional precision is required. The size of the elements then increase away from the bubble, where flow variables vary on a larger spatial scale, so that the problem remains computationally tractable. 

To make clear the arrangement of these smaller elements adjacent to the free surface, Figures~\ref{fig:newmeshb} and \ref{fig:newmeshc} show a close up of the bubble. To accurately capture the dynamics of bubble growth, i.e. the shape of the detaching bubble as well as the pinch--off of the neck, it is found that the mesh adjacent to the neck of the bubble requires more refinement than the mesh above the neck.  

All nodes move along straightened spines apart from those nodes in the pinch--off region. Here, the spines are designed using a bipolar coordinate system with a focus at the contact line. This arrangement is ideal to capture the dynamics of the neck of the bubble as it thins. It was also designed with future research in mind, to capture the dynamics associated with a contact line that is free to move along the solid surface, i.e. the case of a liquid phase that only partially wets the solid surface. In addition, as will be seen later (see Figure~\ref{fig:weep}), under certain conditions, the liquid phase may enter the mouth of the orifice, and this design, in contrast to simpler ones, ensures the free surface can still be captured accurately when this situation arises.   

Each one--dimensional element placed along the free surface consists of three nodes, with the two end nodes being shared with the two neighbouring surface elements. At each time step, the smaller surface elements in the pinch--off region and the larger surface elements around the rest of the bubble are equally spaced, respectively. For all simulations, at least 125 nodes (62 surface elements) were used to track the free surface.  A majority of these nodes, 81, were located in the pinch--off region. In some cases of very large flow rates, simulations used 153 nodes (76 surface elements) due to the increase in the size of the bubble and the necessity to include an additional region of the mesh, within the pinch--off region and adjacent to the contact line, to adequately capture the deformation of the surface that occurs there at the beginning of a simulation. Further increases in the number of nodes is seen to have a negligible impact on the global characteristics of the flow and, remeshing, which can have an adverse effect on accuracy, is not required due to the well--designed meshing strategy.

A simulation runs until the minimum neck radius $r_{min}=r_{tol},$ where the tolerance $r_{tol}= 5\times10^{-2}~r_c.$ Since it is the global characteristics of the bubble that are of interest here and, as stated previously, the problem should remain computationally tractable, more refinement is not added and the simulation stops at this point. As $r_{min} = r_{tol},$ the distance between adjacent nodes on the free surface surrounding the point of minimum neck radius for the largest bubble reported is $5\times10^{-1}$ or $10^{-2}$ for the cases of $r_c=0.1$ and $r_c=1,$ respectively.
 
A second--order backward differentiation formula is used to calculate all derivatives with respect to time. A constant time step is used up until the bubble forms a neck and begins to pinch--off. At least $100$ time steps were used before the neck is formed. This was seen to be small enough so that any decrease in the time step did not sufficiently affect the accuracy of the solution. As the neck thins, around $500$--$700$ time steps are used as the time step decreases logarithmically with the minimum neck radius. As $r_{min} \rightarrow r_{tol},$ the time step is $\mathcal{O}\left(10^{-6}\right)$--$\mathcal{O}\left(10^{-5}\right)$ for $r_c=0.1$ and $\mathcal{O}\left(10^{-4}\right)$--$\mathcal{O}\left(10^{-3}\right)$ for $r_c=1.$
  
Since the free surface of the bubble is the only boundary of the flow domain that is not fixed, as the bubble continues to grow, the size of the flow domain decreases. In order to ensure incompressibility in the flow domain, a `reference' pressure of the liquid must be prescribed at a particular point for the entire simulation. This arbitrary reference pressure should not affect the growth of the bubble and so the straightforward choice would be to base this pressure on the assumption that the pressure field along the side boundary should remain hydrostatic, $p(R,z,t)=-z,$ for the entire simulation. When $p(R,0,t)=0$ is applied as the reference pressure, the point $(R,0)$ behaves like a point sink which affects the velocity field close to the bubble. As this is far from acceptable, $p(R,Z,t)=-Z$ is applied instead. With the values of $R$ and $Z$ stated above, the pressure field along the side boundary remains hydrostatic for the entire solution.

In order to validate the numerical platform, various simulations of a Rayleigh bubble were carried out and the results compared to those obtained from the Rayleigh--Plesset equation \citep{Ples77}. The simulations were seen to converge to the analytic solution with increased spatial and temporal resolution, as expected. 

\section{Results}\label{sec:results}

The influence of the three dimensionless parameters, the orifice radius $r_c,$ the Ohnesorge number $Oh$ and the gas flow rate $Q,$ on the global characteristics of bubble formation, in particular on the dimensionless formation time $t_d$ and volume $V_d,$ can now be investigated. The formation time $t_d$ is taken to be the time of the final solution, as $r_{min}\leq r_{tol},$ whilst the volume $V_d$ of the bubble that is formed is the volume of the bubble above the point on the free surface which represents the minimum neck radius at the formation time. Where the free surface is displayed in a figure, the solid surface will also be shown as a horizontal black line at $r\geq r_c,$ $z=0.$

\subsection{A Typical Case of Bubble Formation}
Figure~\ref{fig:bubble} shows a typical case of bubble formation. Figures (\ref{fig:newexp1})--(\ref{fig:newexp4}) comprise of an experimental image, reproduced from a recent paper of \citet{Bari13}, with the corresponding free surface profile, computed by the numerical platform, superimposed on top. Since Figure~\ref{fig:bubble} shows the entire cross-section of the bubble, the dimensional abscissa is given by $\bar{x}$ rather than $\bar{r}.$ The experimental images show the formation of an air bubble in water ($\rho=998.2$~kg~m$^{-3}$, $\mu=10^{-3}$~Pa~s, $\sigma=73$~mN~m$^{-1}$) by applying a dimensional volumetric gas flow rate of $\bar{Q}=2.78$~mm$^3$~s$^{-1}$ through an orifice of dimensional radius $\bar{r}_c=0.29$~mm in a submerged solid surface. There is very good agreement between the simulation and the experiment.

With $g=9.81$~m~s$^{-2}$ and using the scales identified in the problem formulation, the material parameters of the liquid give rise to the length scale $L=2.73$~mm, velocity scale $U=73$~m~s$^{-1},$ pressure scale $\sigma/L = 26.7$~Pa, time scale $L/U = 3.74\times10^{-5}$~s and flow rate scale $L^2 U=5.44\times10^{-4}$~m$^3$~s$^{-1}.$ The experiment is then classified as the dimensionless case of $(r_c,Oh,Q)=(0.106,2.24\times10^{-3},5.11\times10^{-6}).$ 

The dimensional experimental times $\bar{t}_e$ that accompany each image are $9.60\times10^{-2}$~s, $1.380$~s, $2.340$~s and $3.550$~s, respectively. However, the initial configuration at $\bar{t}_e=0$ is unclear as that result was not published and so, in order to compare the simulation to the experiment, the simulated free surface profile that best matched the fourth experimental image (see Figure~\ref{fig:newexp4}) was selected. This profile occurred at the dimensional simulation time $\bar{t}=4.137$~s, and so the criterion $\bar{t}=\bar{t}_e+0.587$ was used to find the corresponding simulation time of the first three experimental images. The dimensional simulation times of the four computed solutions are therefore $\bar{t}=0.683$~s, $1.967$~s, $2.927$~s and $4.137$~s. 

\begin{figure}[t]
\begin{center}
\includegraphics[width=7.9cm,keepaspectratio]{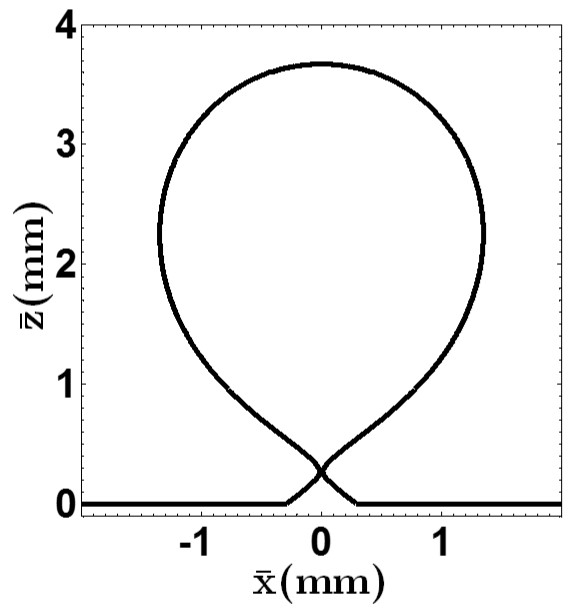}
\caption{The final computed solution of the case $(r_c,Oh,Q)=(0.106,2.24\times10^{-3},5.11\times10^{-6}),$ as $r_{min}=r_{tol},$ where $\bar{t}=4.139$~s and so $t_d=1.107\times10^5.$}
\label{fig:newexp5}
\end{center}
\end{figure}
At $\bar{t}=0$~s, the simulated bubble starts as a spherical cap of volume $1.677\times10^{-2}$~mm$^3.$ Initially, the bubble grows spherically as the force of capillarity controls the early growth (see Figure~\ref{fig:newexp1} and \ref{fig:newexp2}). The volume of the bubble at $\bar{t}=0.683$~s and $1.967$~s is $1.937$~mm$^3$ and $5.473$~mm$^3$, respectively. By $\bar{t}=2.927$~s, the force of buoyancy becomes important on the bubble of volume $8.176$~mm$^3.$ The bubble translates upwards and begins to lose sphericity above the contact line (see Figure~\ref{fig:newexp3}).

As the bubble seeks to minimise its surface area at a given volume and due to the fact that the hydrostatic pressure varies linearly along the bubble, a neck begins to form in the free surface. This can be seen at $\bar{t}=4.137$~s (see Figure~\ref{fig:newexp4}) where the volume of the bubble is now $11.518$~mm$^3$. Once the neck has formed, as there is a change in the sign of the longitudinal curvature at some point on the free surface, the pinch--off process can begin. The increasing difference in hydrostatic pressure between the apex and the base of the bubble results in an increase of capillary pressure at the neck which then drives the shrinking of the neck still further.

Figure~\ref{fig:newexp5} shows the final computed solution, as $r_{min}=r_{tol},$ at $\bar{t}=4.139$~s. Here, the total volume of the bubble is $11.523$~mm$^3,$ whilst the volume of the newly formed bubble, measured above the point of minimum neck radius is $11.502$~mm$^3.$ Therefore, this final solution corresponds to a dimensionless formation time of $t_d=1.107\times10^5$ and a dimensionless bubble volume of $V_d=0.5651.$ 

Compared to the overall generation of the bubble, which took $4.139$~s, the pinch--off process takes place much faster in $1.68\times10^{-2}$~s. Notably, while the experiments appear unable to capture the very small time scales in the final stages of pinch--off, this is no barrier for the computations.

\begin{figure}[t]
\begin{center}
\includegraphics[width=0.8\textwidth]{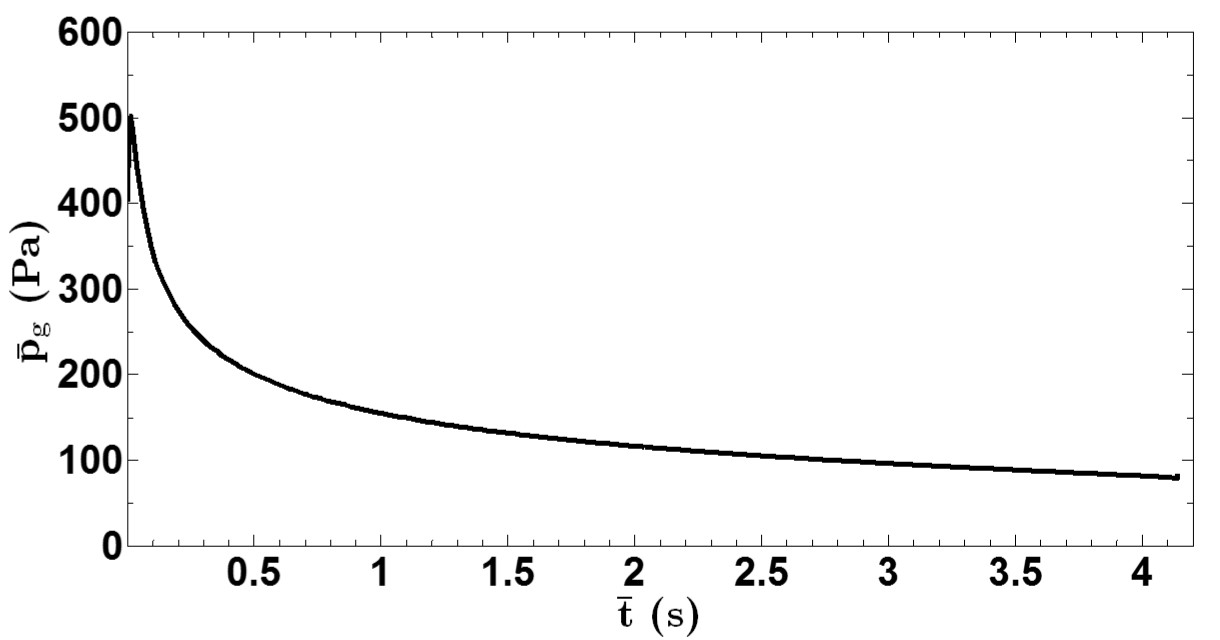}
\caption{The evolution of the dimensional gas pressure $\bar{p}_g$ with dimensional simulation time $\bar{t}$ for the case of $(r_c,Oh,Q)=(0.106,2.24\times10^{-3},5.11\times10^{-6}).$}
\label{fig:newgaspres}
\end{center}
\end{figure} 
The evolution of the dimensional gas pressure during the simulation is shown by Figure~\ref{fig:newgaspres}. Recalling that the pressure field along the side boundary of the domain remains hydrostatic with $p(R,0,t)=0,$ the gas pressure reaches its greatest value near the start of the simulation as the bubble resembles a hemisphere. The gas pressure then decreases by almost $80\%$ until there is a very slight increase as $r_{min}\rightarrow r_{tol},$ which can only be seen under the appropriate magnification. 

\subsection{Influence of Parameters}
\begin{figure}[]
\begin{center}
	\begin{subfigure}[]{0.85\textwidth}
	\centering
	\includegraphics[width=14cm,keepaspectratio]{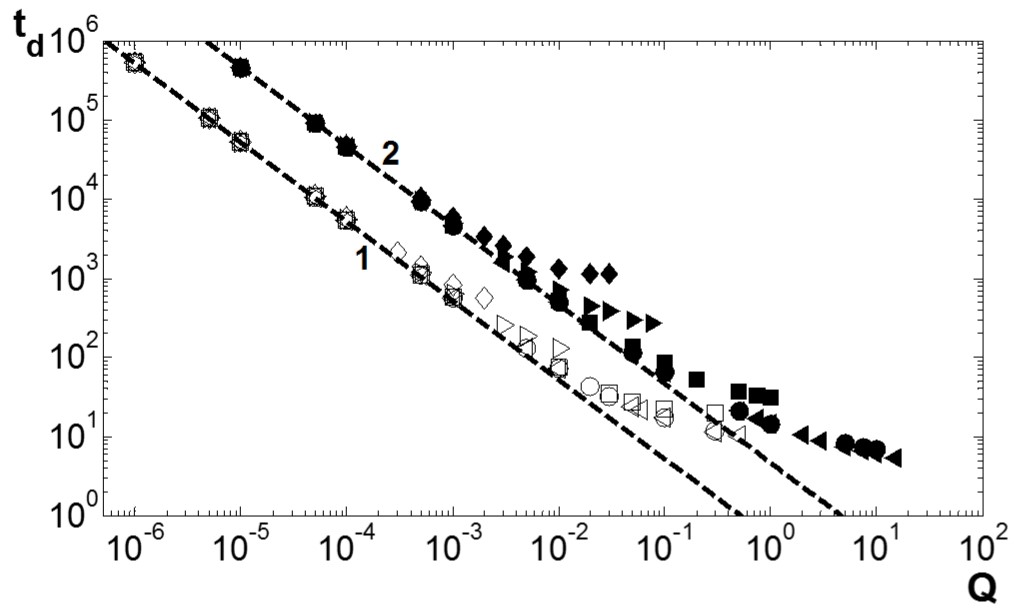}
	\caption{A log--log plot of the dimensionless formation time $t_d$ of the bubble versus the dimensionless flow rate $Q$ for a variety of Ohnesorge numbers $Oh$. Line $1$ is $t_d=0.53/Q$ and line $2$ is $t_d=4.7/Q.$}
	\label{fig:TvQ}
	\end{subfigure}
	
	\begin{subfigure}[]{0.85\textwidth}
	\centering
	\includegraphics[width=14cm,keepaspectratio]{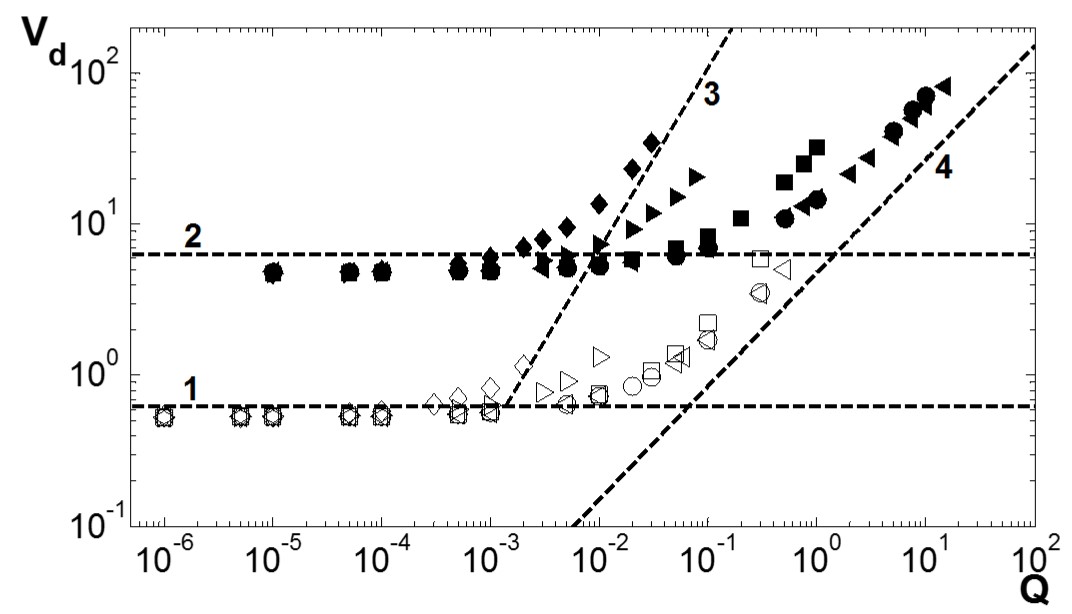}
	\caption{A log--log plot of the dimensionless bubble volume $V_d$ versus the dimensionless flow rate $Q$ for a variety of Ohnesorge numbers $Oh$. Line $1$ is $V_F=6.28 \times 10^{-1},$ line $2$ is $V_F=6.28,$ line $3$ is equation (\ref{eq:oguz}) with $Oh=2.24\times10^{-3}$ and line $4$ is equation (\ref{eq:wong}).}
	\label{fig:VvQ}
	\end{subfigure}
	\caption{A map of the parameter space where open symbols correspond to $r_c=0.1$ and filled symbols correspond to $r_c=1.$ The various Ohnesorge numbers are represented by ($\diamond$) $Oh=2.24\times10^{-3},$ ($\triangleright$) $Oh=10^{-2},$ ($\square$) $Oh=10^{-1},$ ($\circ$) $Oh=1$ and ($\triangleleft$) $Oh=10.$ } 
	\label{fig:map}
\end{center}
\end{figure}
Figure~\ref{fig:map} shows how the global characteristics of bubble formation, the dimensionless formation time $t_d$ (see Figure~\ref{fig:TvQ}) and the dimensionless bubble volume $V_d$ (see Figure~\ref{fig:VvQ}), depend upon the gas flow rate $Q$ and the Ohnesorge number $Oh$ for the dimensionless orifice radii of $r_c=0.1$ and $1.$ These benchmark calculations fall into two regimes, the regimes of low gas flow rates and high gas flow rates, otherwise known as the `static' and `dynamic' regimes, respectively. These regimes are now considered separately. 

\subsubsection{Regime of Low Gas Flow Rates}
For a relatively small gas flow rate, a variation in the Ohnesorge number has very little effect on the formation time (see Figure~\ref{fig:TvQ}) and, consequently, very little effect on the bubble volume (see Figure~\ref{fig:VvQ}). Therefore, for a given orifice radius, as $Q\rightarrow0,$ then $V_d\rightarrow V_c,$ where $V_c$ is the limiting bubble volume. Specifically, $V_c=0.53$ and $4.7$ for $r_c=0.1$ and $1.0,$ respectively. In other words, for a given orifice radius, simply decreasing the flow rate does not lead to smaller bubble volumes in this regime. The force of buoyancy is simply not large enough to detach the bubble from the formation site until the bubble volume reaches $V_c.$ 

By assuming that the bubble remains spherical and balancing the forces of buoyancy and capillarity, \citet{Fritz35} (see \citet{Kuma70}) derived an expression for this limiting bubble volume which, when rescaled using the scales found in the problem formulation, is given by, 
\begin{equation} V_F=2\pi r_c.\end{equation}
Therefore $V_F=6.28 \times 10^{-1}$ and $6.28$ for $r_c=0.1$ and $1,$ respectively (see lines $1$ and $2$ respectively on Figure~\ref{fig:VvQ}). This gives way to large relative errors of $18.5\%$ and $33.6\%$ for $r_c=0.1$ and $1,$ respectively, when compared to the simulated limiting volume $V_c.$ 

Since the volume of the newly formed bubble is much greater than the volume of the residual bubble then $t_d \approx V_c/Q$ in this regime. Therefore, for $r_c=0.1$ and $1,$ the respective formation times can be expressed as $t_d=0.53/Q$ and $t_d=4.7/Q$ when $Q<10^{-4}$ (see lines $1$ and $2$ respectively in Figure~\ref{fig:TvQ}). 

As expected, the limiting bubble volume $V_c$ increases with orifice radius. These conclusions are in agreement with the literature where the regime of low gas flow rates is also known as the `static' regime. As will now be shown below, much of the growth of the bubble can be described by a quasi--static approach that involves the Young--Laplace equation.

\begin{figure}[t]
\begin{center}
	\begin{subfigure}[]{0.45\textwidth}
	\centering
	\includegraphics[width=7cm,keepaspectratio]{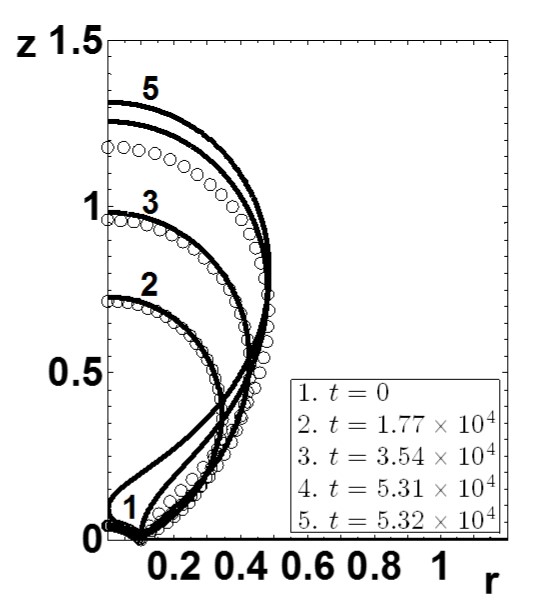}
	\caption{The case of $(r_c,Oh,Q)=(0.1,1,10^{-5}).$}
	\label{fig:static01}
	\end{subfigure}
	\qquad
	\begin{subfigure}[]{0.45\textwidth}
	\centering
	\includegraphics[width=6cm,keepaspectratio]{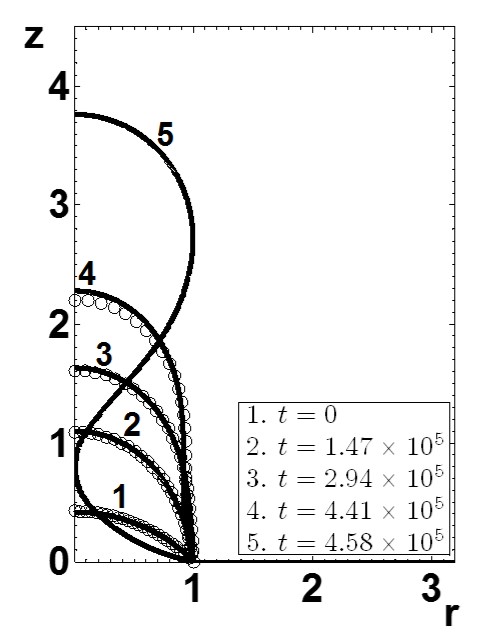}
	\caption{The case of $(r_c,Oh,Q)=(1,1,10^{-5}).$}
	\label{fig:static1}
	\end{subfigure}
	\caption{The formation of bubbles in the regime of low gas flow rates. The dynamic finite element simulations are given by the black lines whilst the circular symbols represent the finite difference Young--Laplace solutions for quasi--static growth.}
	\label{fig:static}
\end{center}
\end{figure}
Figure~\ref{fig:static} shows the simulated temporal evolution of a bubble (continuous black lines) for two cases in the regime of low gas flow rates, $(r_c, Oh, Q)=(0.1, 1, 10^{-5})$ and $(1,1,10^{-5}),$ respectively. In both cases, curve $1$ is the initial solution and curve $4$ shows the free surface when the neck first begins to develop. Curves $2$ and $3$ are equally spaced in time between curves $1$ and $4,$ where it can be seen that, once again, the capillarity force dominates the early stages of bubble growth. This is particularly evident in the case of $r_c=0.1$ (see Figure~\ref{fig:static01}) where the bubble grows spherically for almost the entire formation time due to its smaller size than the case of $r_c=1.0$ (see Figure~\ref{fig:static1}). As the pinch--off process continues, the bubble approaches break up and the final solution, as $r_{min}=r_{tol},$ is given by curve $5$ where a bubble of volume $V_d=0.532$ or $4.90$ are formed for $r_c=0.1$ or $1,$ respectively.

Due to the relatively small liquid velocities associated with the initial bubble growth in these cases, the initial growth can be accurately described by a series of quasi--static profiles that are found by solving the Young--Laplace equation and stipulating a small increase in volume from one solution to the next. 

Assuming that liquid velocities are negligible and rescaling, the dimensionless Young--Laplace equation is,
\begin{equation}\label{eq:YL} p_g+z=\kappa_1 + \kappa_2, \end{equation}   
where $\kappa_1$ and $\kappa_2$ are the respective dimensionless longitudinal and cross--sectional curvatures at each point on the free surface and the liquid pressure is assumed to be hydrostatic $p(r,z,t)=-z.$ 

\citet{Ford48} and \citet{Thor05} described a scheme that can be used to solve (\ref{eq:YL}), which involves a system of first--order ordinary differential equations. The finite difference method can be used to solve this system and so for a given finite element profile, the corresponding quasi--static solution, with the same bubble volume as the simulation, can then be found. The uniform gas pressure in (\ref{eq:YL}) is found as part of each quasi-static solution.

These series of quasi--static profiles are also shown in Figure~\ref{fig:static} with the circular symbols. There is very good agreement between the first three quasi--static and finite element solutions. However, as the bubble continues to grow, the quasi--static solutions become less accurate, as seen by the fourth solutions. Any further quasi--static solutions are wildly inaccurate. 

In summary, in the regime of low gas flow rates, the initial growth of the bubble is a quasi--static process that is governed by the gas, hydrostatic and capillary pressures. As the neck forms and pinch--off begins, bubble growth is an essentially dynamic process, and unsurprisingly, the quasi--static approach using the Young--Laplace equation (\ref{eq:YL}) is inadequate in describing the effects of the large velocities associated with the thinning of the neck together with the importance of inertia and viscosity in the liquid. 

\subsubsection{Regime of High Gas Flow Rates}
Greater gas flow rates than those associated with the previous regime are now considered. For a given Ohnesorge number, as the flow rate is increased, it is now the formation time that tends to a limiting value $t_c$ (see Figure~\ref{fig:TvQ}). In other words, when the flow rate is sufficiently large, a bubble can not form quicker than this limiting time. This effect is most clear in the case of $r_c=1.$ As $Q$ increases, the difference between the formation time of the simulations and the formation time associated with the regime of low gas flow rates ($t_d=0.53/Q$ and $4.7/Q$ for $r_c=0.1$ and $1,$ respectively) increases. In agreement with the literature, this results in an increase in bubble volume $V_d$ with flow rate (see Figure~\ref{fig:VvQ}).

For a given gas flow rate, the formation time increases with decreasing Ohnesorge number and therefore, due to the bubble inflating at a constant volumetric flow rate, the bubble volume increases with decreasing Ohnesorge number. This is once again more obvious in the case of $r_c=1.0$ as the formation times tend to a limiting value. As the Ohnesorge number increases, the limiting formation time decreases. For $r_c=1,$ $t_c \approx 1130,$ $260,$ $30,$ $7$ and $5$ for $Oh=2.24\times10^{-3},$ $10^{-2},$ $10^{-1},$ $1$ and $10,$ respectively (see Figure~\ref{fig:TvQ}). The reason for this is the increased inertia in the liquid opposes the pinching of the neck which leads to a prolonging of the formation time. There is very little difference in the results between an Ohnesorge number of $1$ and $10,$ suggesting that inertial effects become negligible for $Oh>1.$ Once again, it is worth reiterating that the bubbles formed at a given flow rate from a larger orifice have a larger formation time and therefore a greater volume.

\citet{Oguz93} used a two--stage model which includes the forces of inertia, capillarity and buoyancy to derive a critical flow rate at which, in the case of an inviscid liquid, bubble formation transitions from the low flow rate regime to the high flow rate regime. By rescaling, this critical flow rate is given by, 
\begin{equation}\label{eq:oguzq} Q_{cr}=\left(16/3\right)^{1/6} \pi~Oh~r_c^{5/6}.\end{equation}
For a flow rate $Q \leq Q_{cr},$ $V_d=V_F,$ whilst for $Q>Q_{cr},$ it was found that,
\begin{equation}\label{eq:oguz} V_0=A_1\left(\frac{Q}{Oh}\right)^{6/5}, \,\,\,\,\,\,A_1=\frac{4 \pi}{3}\left(\frac{9}{8\pi^2}\right)^{3/5},\end{equation} 
where the subscript $0$ represents the inviscid regime, $Oh\rightarrow0.$ 

To compare (\ref{eq:oguz}) with the results presented here, line $3$ in Figure~\ref{fig:VvQ} is (\ref{eq:oguz}) with $Oh=2.24\times10^{-3},$ the smallest Ohnesorge number simulated. The simulated bubble volumes for $Oh=2.24\times10^{-3}$ and $r_c=1.0$ approach those values predicted by (\ref{eq:oguz}) as the flow rate increases. It is unclear whether the scaling law is valid for $r_c=0.1$ because the greater flow rates required could not be computed by the numerical platform.

Once again, the case of $Oh=2.24\times10^{-3}$ is used to examine the critical flow rate given by (\ref{eq:oguzq}), where $Q_{cr}=1.37\times10^{-3}$ and $9.3\times10^{-3}$ for $r_c=0.1$ and $1,$ respectively. These flow rates are also given by the points of intersection of line $3$ by line $1$ for $r_c=0.1$ and line $2$ for $r_c=1$ in Figure~\ref{fig:VvQ}. It can be seen from the simulations that (\ref{eq:oguzq}) overestimates the flow rate at which inviscid bubble formation transitions from the regime of low flow rates to the regime of high flow rates. 

In the case of a highly viscous liquid, \citet{Wong98} used a spherical model and balanced the viscous and capillarity forces to derive an expression for the bubble volume where $r_c \rightarrow 0.$ When rescaled,
\begin{equation}\label{eq:wong} V_{\infty}=A_2 Q^{3/4}, \,\,\,\,\,\,A_2=\left(500 \pi/3\right)^{1/4}, \end{equation}
where the subscript $\infty$ represents the highly viscous regime of $Oh\rightarrow\infty.$ The volumes predicted by (\ref{eq:wong}) underestimate those given by the simulations for the most viscous case of $Oh=10$ for both $r_c=0.1$ and $1$ (see line $4$ on Figure~\ref{fig:VvQ}). 

It is possible to interpolate between the scaling laws (\ref{eq:oguz}) and (\ref{eq:wong}), for the respective limiting cases of $Oh\rightarrow0$ and $Oh\rightarrow\infty,$ to fit a scaling law to the results for intermediate Ohnesorge numbers shown in Figure~\ref{fig:VvQ}. The scaling law for bubble volume is assumed to take the form,
\begin{equation}\label{eq:scale} V_d=\frac{Q^{6/5+c~f(Oh)}}{f(Oh)},\,\,\,\,\,\,f(Oh)=\frac{a~Oh^{6/5}}{b~Oh^{6/5}+1}.\end{equation}
Then, applying $V_d \rightarrow V_0$ as $Oh \rightarrow 0,$ and $V_d \rightarrow V_{\infty}$ as $Oh \rightarrow \infty,$ gives, $$a=1/A_1,\,\,\,\, b=A_2/A_1,\,\,\,\, c=-9~A_2/20,$$ 
where $V_0$ and $A_1$ are given by (\ref{eq:oguz}) and $V_{\infty}$ and $A_2$ are given by (\ref{eq:wong}). 

\begin{figure}[t]
\begin{center}
\includegraphics[width=0.85\textwidth]{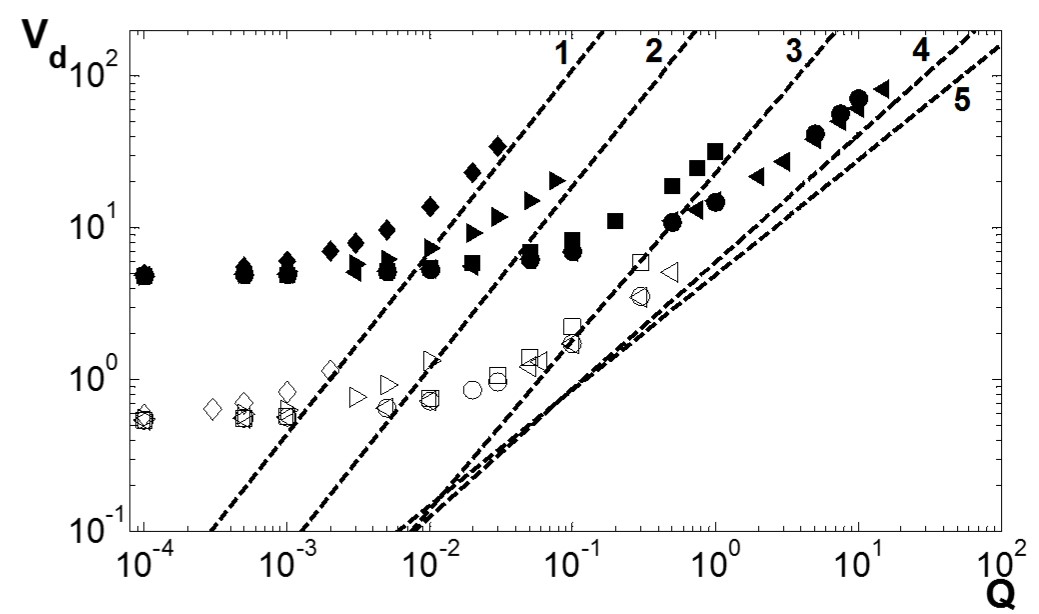}
\caption{Comparing the scaling law given by equation~(\ref{eq:scale}) for $V_d$ in the high gas flow rate regime, depicted by the dashed lines, against the results presented in Figure~\ref{fig:VvQ} for $Oh=2.24\times10^{-3}$ (line $1$), $Oh=10^{-2}$ ($2$), $Oh=10^{-1}$, ($3$), $Oh=1,$ ($4$) and $Oh=10$ ($5$).}
\label{fig:scaling}
\end{center}
\end{figure} 
For each Ohnesorge number investigated in this work, Figure~\ref{fig:scaling} shows that as the flow rate $Q$ increases, the results from the simulations tend towards those predicted by the scaling law (\ref{eq:scale}) as $Q \rightarrow \infty,$ but do not give an accurate representation for moderate $Q.$ Therefore, whilst scaling laws may give a valid approximation of the global characteristics of bubble formation in certain regimes, simulations are required to obtain a more accurate representation of the entire parameter space. 

\begin{figure}[]
\begin{center}
	\begin{subfigure}[]{0.22\textwidth}
	\centering
	\includegraphics[width=4cm,keepaspectratio]{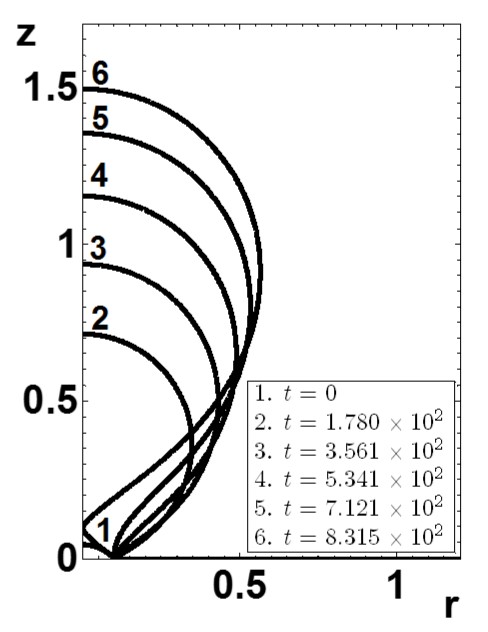}
	\caption{$Oh=2.24\times10^{-3},$\\ $Q=10^{-3}.$}
	\label{fig:818}
	\end{subfigure}
	\qquad
	\begin{subfigure}[]{0.19\textwidth}
	\centering
	\includegraphics[width=4cm,keepaspectratio]{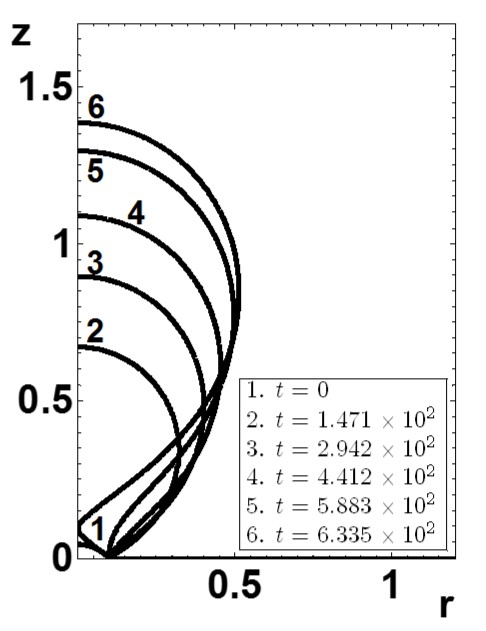}
	\caption{$Oh=10^{-2},$\\ $Q=10^{-3}.$}
	\label{fig:815}
	\end{subfigure}
	\qquad
	\begin{subfigure}[]{0.19\textwidth}
	\centering
	\includegraphics[width=4cm,keepaspectratio]{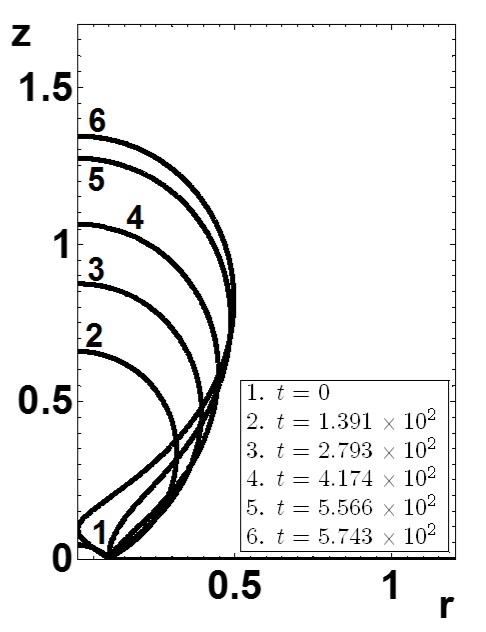}
	\caption{$Oh=10^{-1},$\\ $Q=10^{-3}.$}
	\label{fig:812}
	\end{subfigure}
	\qquad
	\begin{subfigure}[]{0.19\textwidth}
	\centering
	\includegraphics[width=4cm,keepaspectratio]{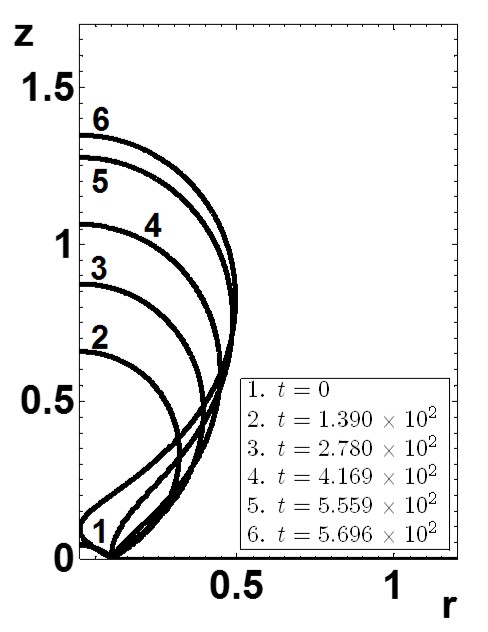}
	\caption{$Oh=1,$\\ $Q=10^{-3}.$}
	\label{fig:809}
	\end{subfigure}
	
	\begin{subfigure}[]{0.22\textwidth}
	\centering
	\includegraphics[width=4cm,keepaspectratio]{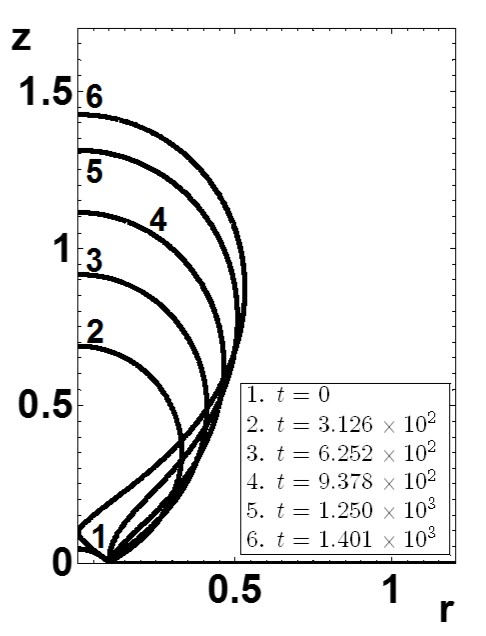}
	\caption{$Oh=2.24\times10^{-3},$\\ $Q=5\times10^{-4}.$}
	\label{fig:817}
	\end{subfigure}
	\qquad
	\begin{subfigure}[]{0.19\textwidth}
	\centering
	\includegraphics[width=4cm,keepaspectratio]{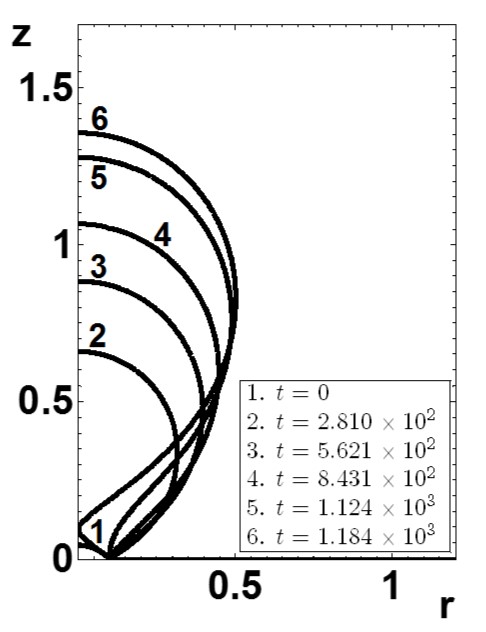}
	\caption{$Oh=10^{-2},$\\ $Q=5\times10^{-4}.$}
	\label{fig:814}
	\end{subfigure}
	\qquad
	\begin{subfigure}[]{0.19\textwidth}
	\centering
	\includegraphics[width=4cm,keepaspectratio]{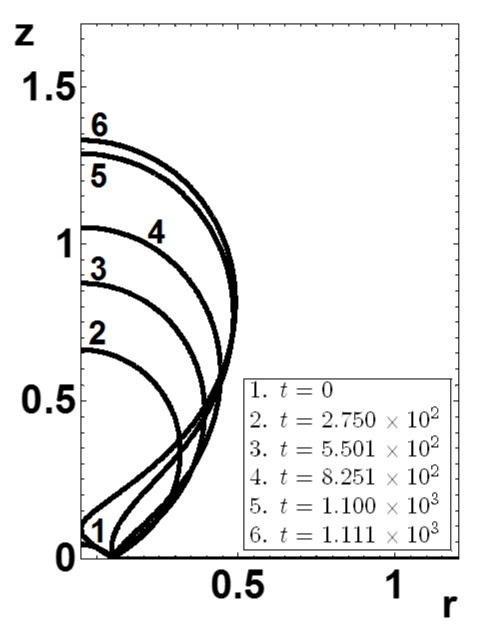}
	\caption{$Oh=10^{-1},$\\ $Q=5\times10^{-4}.$}
	\label{fig:811}
	\end{subfigure}
	\qquad
	\begin{subfigure}[]{0.19\textwidth}
	\centering
	\includegraphics[width=4cm,keepaspectratio]{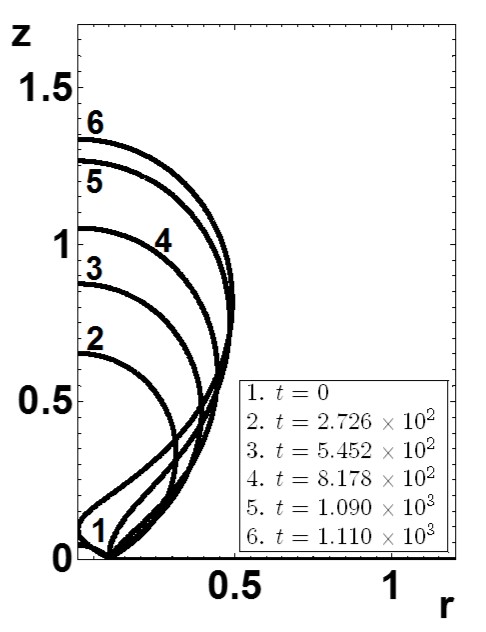}
	\caption{$Oh=1,$\\ $Q=5\times10^{-4}.$}
	\label{fig:808}
	\end{subfigure}

	\begin{subfigure}[]{0.22\textwidth}
	\centering
	\includegraphics[width=4cm,keepaspectratio]{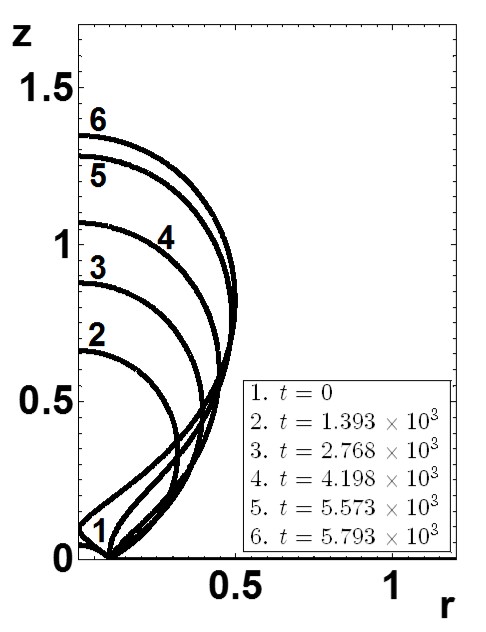}
	\caption{$Oh=2.24\times10^{-3},$\\ $Q=10^{-4}.$}
	\label{fig:816}
	\end{subfigure}
	\qquad
	\begin{subfigure}[]{0.19\textwidth}
	\centering
	\includegraphics[width=4cm,keepaspectratio]{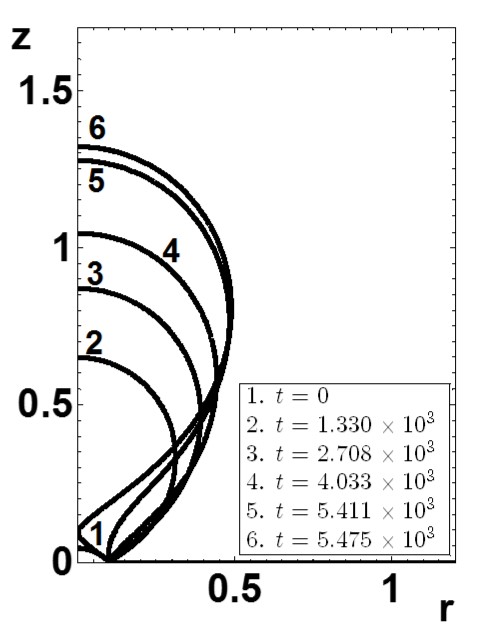}
	\caption{$Oh=10^{-2},$\\ $Q=10^{-4}.$}
	\label{fig:813}
	\end{subfigure}
	\qquad
	\begin{subfigure}[]{0.19\textwidth}
	\centering
	\includegraphics[width=4cm,keepaspectratio]{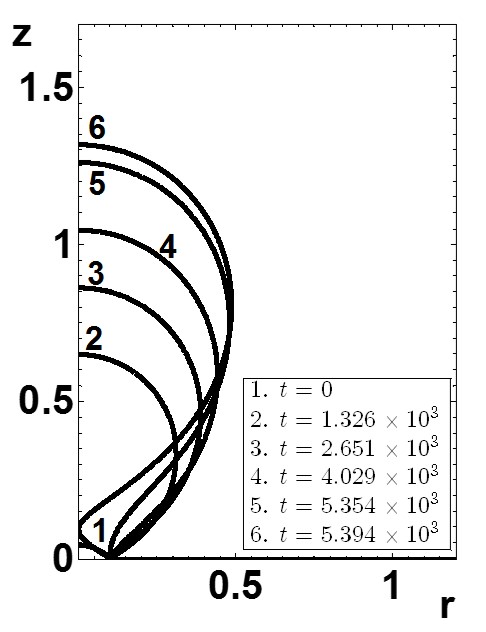}
	\caption{$Oh=10^{-1},$\\ $Q=10^{-4}.$}
	\label{fig:810}
	\end{subfigure}
	\qquad
	\begin{subfigure}[]{0.19\textwidth}
	\centering
	\includegraphics[width=4cm,keepaspectratio]{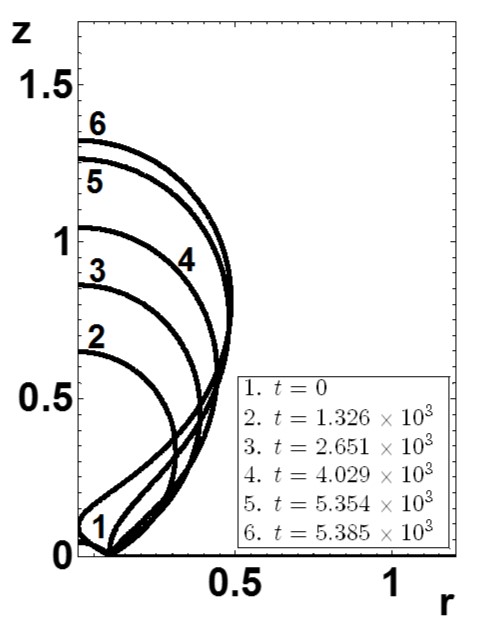}
	\caption{$Oh=1,$\\ $Q=10^{-4}.$}
	\label{fig:807}
	\end{subfigure}
	\caption{Temporal evolution of the free surface for various Ohnesorge numbers $Oh$ and flow rates $Q$ for an orifice of radius $r_c=0.1.$ }
	\label{fig:profile01}
\end{center}
\end{figure} 

\begin{figure}[]
\begin{center}
	\begin{subfigure}[]{0.22\textwidth}
	\centering
	\includegraphics[width=4cm,keepaspectratio]{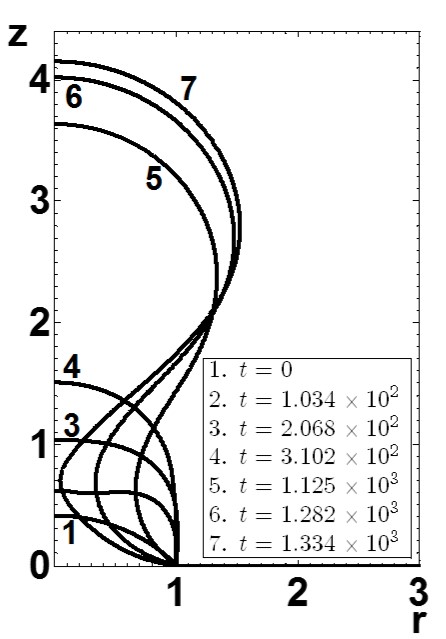}
	\caption{$Oh=2.24\times10^{-3},$\\ $Q=10^{-2}.$}
	\label{fig:1005}
	\end{subfigure}
	\qquad
	\begin{subfigure}[]{0.19\textwidth}
	\centering
	\includegraphics[width=4cm,keepaspectratio]{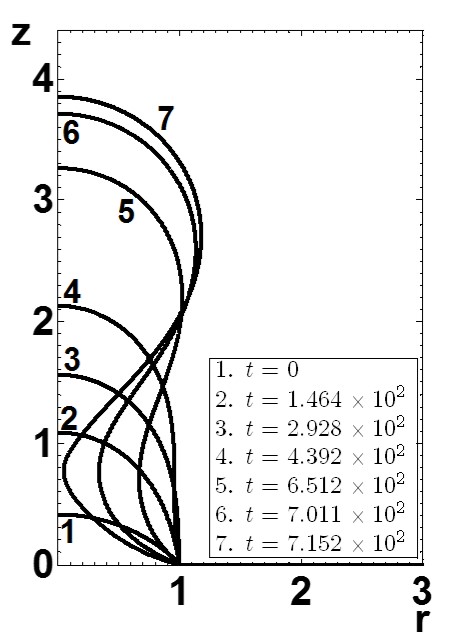}
	\caption{$Oh=10^{-2},$\\ $Q=10^{-2}.$}
	\label{fig:988}
	\end{subfigure}
	\qquad
	\begin{subfigure}[]{0.19\textwidth}
	\centering
	\includegraphics[width=4cm,keepaspectratio]{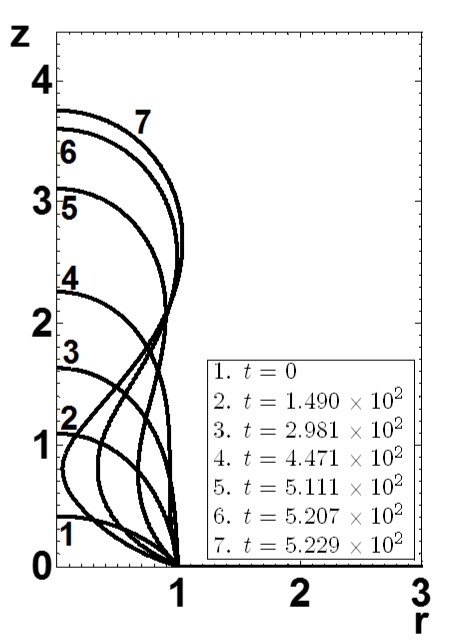}
	\caption{$Oh=10^{-1},$\\ $Q=10^{-2}.$}
	\label{fig:984}
	\end{subfigure}
	\qquad
	\begin{subfigure}[]{0.19\textwidth}
	\centering
	\includegraphics[width=4cm,keepaspectratio]{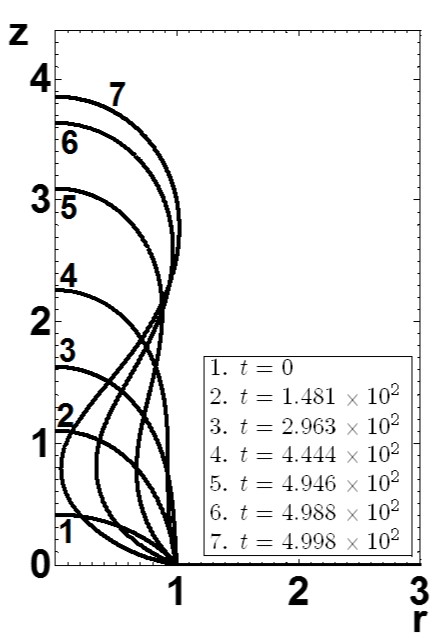}
	\caption{$Oh=1,$\\ $Q=10^{-2}.$}
	\label{fig:820}
	\end{subfigure}
	
	\begin{subfigure}[]{0.22\textwidth}
	\centering
	\includegraphics[width=4cm,keepaspectratio]{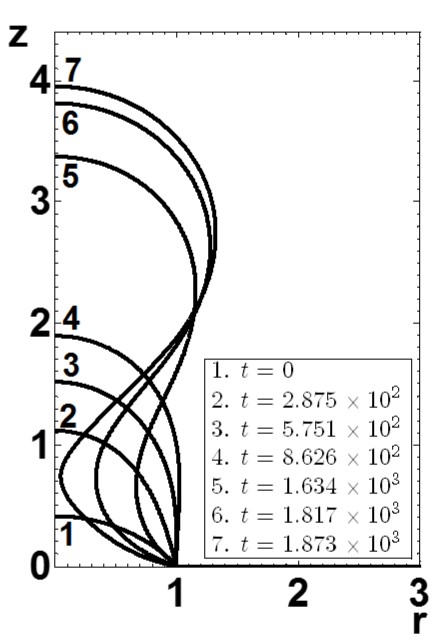}
	\caption{$Oh=2.24\times10^{-3},$\\ $Q=5\times10^{-3}.$}
	\label{fig:945}
	\end{subfigure}
	\qquad
	\begin{subfigure}[]{0.19\textwidth}
	\centering
	\includegraphics[width=4cm,keepaspectratio]{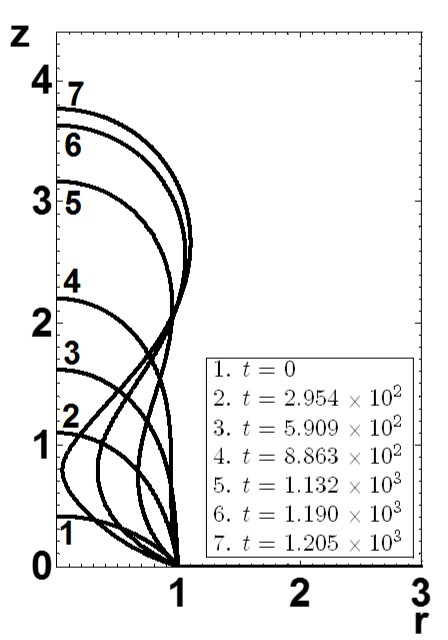}
	\caption{$Oh=10^{-2},$\\ $Q=5\times10^{-3}.$}
	\label{fig:975}
	\end{subfigure}
	\qquad
	\begin{subfigure}[]{0.19\textwidth}
	\centering
	\includegraphics[width=4cm,keepaspectratio]{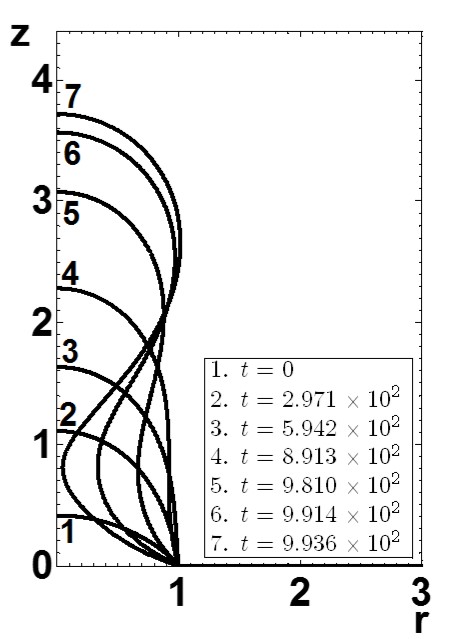}
	\caption{$Oh=10^{-1},$\\ $Q=5\times10^{-3}.$}
	\label{fig:983}
	\end{subfigure}
	\qquad
	\begin{subfigure}[]{0.19\textwidth}
	\centering
	\includegraphics[width=4cm,keepaspectratio]{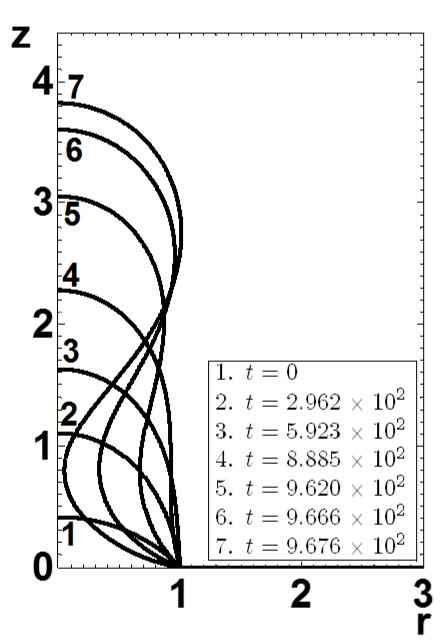}
	\caption{$Oh=1,$\\ $Q=5\times10^{-3}.$}
	\label{fig:825}
	\end{subfigure}

	\begin{subfigure}[]{0.22\textwidth}
	\centering
	\includegraphics[width=4cm,keepaspectratio]{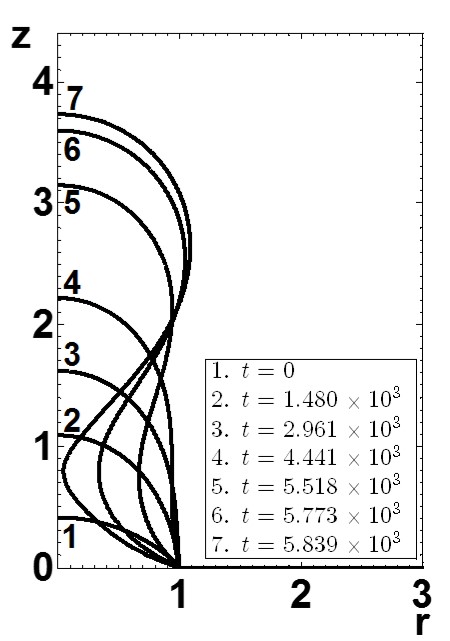}
	\caption{$Oh=2.24\times10^{-3},$\\ $Q=10^{-3}.$}
	\label{fig:955}
	\end{subfigure}
	\qquad
	\begin{subfigure}[]{0.19\textwidth}
	\centering
	\includegraphics[width=4cm,keepaspectratio]{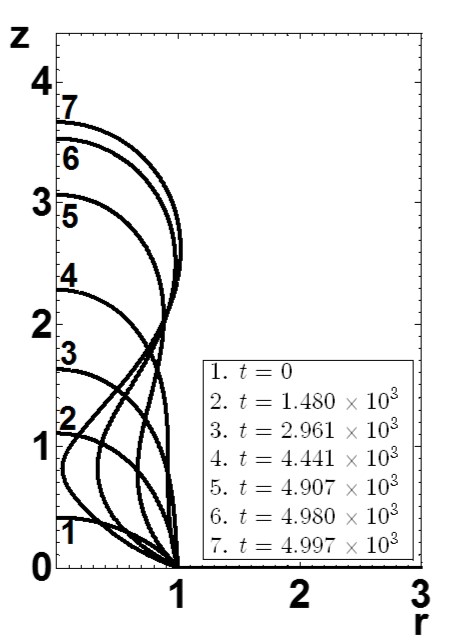}
	\caption{$Oh=10^{-2},$\\ $Q=10^{-3}.$}
	\label{fig:954}
	\end{subfigure}
	\qquad
	\begin{subfigure}[]{0.19\textwidth}
	\centering
	\includegraphics[width=4cm,keepaspectratio]{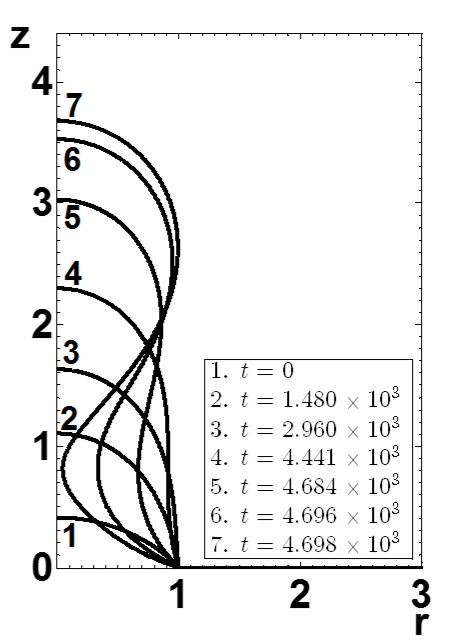}
	\caption{$Oh=10^{-1},$\\ $Q=10^{-3}.$}
	\label{fig:953}
	\end{subfigure}
	\qquad
	\begin{subfigure}[]{0.19\textwidth}
	\centering
	\includegraphics[width=4cm,keepaspectratio]{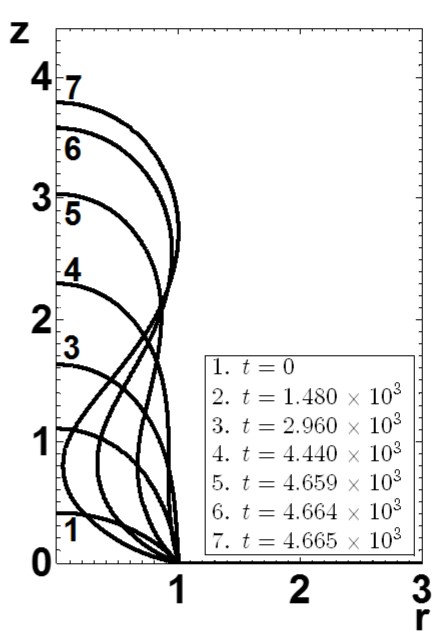}
	\caption{$Oh=1,$\\ $Q=10^{-3}.$}
	\label{fig:952}
	\end{subfigure}
	\caption{Temporal evolution of the free surface for various Ohnesorge numbers $Oh$ and flow rates $Q$ for an orifice of radius $r_c=1.0.$ }
	\label{fig:profile1}
\end{center}
\end{figure} 
Figures~\ref{fig:profile01} and \ref{fig:profile1} show the evolution of the bubble for a cross--section of the parameter space in the regime of high gas flow rates for $r_c=0.1$ and $1.0$, respectively. As discussed previously, the cases of $Oh=1$ and $Oh=10$ are very similar and so the case of $Oh=10$ is not given in Figures~\ref{fig:profile01} and \ref{fig:profile1}.  

The plots in Figure~\ref{fig:profile01} each show six curves. Curve $1$ is the initial solution, curve $5$ is where the bubble first obtains a neck and curve $6$ is the final solution as $r_{min} = r_{tol}.$ Curves $2,$ $3$ and $4$ are equally spaced in time between curves $1$ and $5.$ The plots in Figure~\ref{fig:profile1} have an extra curve. Here, curve $4$ is where the neck begins to develop and curve $7$ is the final solution, curves $2$ and $3$ are equally spaced in time between curves $1$ and $4,$ and curves $5$ and $6$ show the bubble when $r_{min}=2/3~r_c$ and $r_{min}=1/3~r_c,$ respectively.

Due to the greater velocity imparted in the liquid from the larger flow rates and because inertia cannot be ignored at these flow rates, the aforementioned quasi--static approach, using the Young--Laplace equation (\ref{eq:YL}), to compute free surface profiles, is valid for a much smaller proportion of the overall formation process than was seen in Figure~\ref{fig:static} for bubble formation under low gas flow rates. For greater flow rates, the bubble formation process is almost entirely dynamic and therefore this regime of high gas flow rates is also known as the `dynamic' regime.

A major difference between bubbles forming from orifices of different radii is the increased sphericity of those bubbles for $r_c=0.1$ due to the larger surface--to--volume ratio where surface effects are more important for smaller systems. Also the bubbles generated in the case of $r_c=0.1$ are much larger relative to the orifice radius than those in the case of $r_c=1.0,$ which is to be expected since $r_c=1.0$ corresponds to the dimensional case where the orifice radius is equal to the length scale. 

Let $t_d=t_g+t_n,$ where $t_g$ is the time taken by the initial growth stage of the bubble, until the longitudinal curvature of the free surface changes sign at a point as the neck forms, and $t_n$ is the pinch--off time, taken to be the time taken for $r_{min}= r_{tol}$ once the neck has formed. 

For a given flow rate, the ratio $t_n/t_d$ increases with decreasing Ohnesorge number. The influence of inertia in prolonging the formation time by opposing the thinning of the neck of the bubble is illustrated by Figures~\ref{fig:807} and \ref{fig:810} for $r_c=0.1,$ and by Figures~\ref{fig:952} and \ref{fig:955} for $r_c=1.0,$ where the times taken for the necks to form are almost equal, yet $t_d$ increases with decreasing Ohnesorge number. This is further highlighted by the rapid vertical displacement of the apex of the bubble seen in Figure~\ref{fig:820}, where, in order to keep the flow rate constant at $Q=10^{-2},$ the apex must rise quickly to make up the volume lost due to this rapid pinch--off. 

For a given Ohnesorge number, as the flow rate increases, the majority of the increase in volume is accounted for by the increased width of the bubbles, see for example, Figures~\ref{fig:816} and \ref{fig:818} for $r_c=0.1$, and Figures~\ref{fig:955} and \ref{fig:1005}, for $r_c=1.$ The ratio $t_n/t_d$ initially increases with increasing flow rate as seen in Figure~\ref{fig:profile01} and \ref{fig:profile1}. As the flow rate continues to increase $t_n/t_d$ reaches a maximum before decreasing with increasing flow rate.  

\begin{figure}[t]
\begin{center}
	\begin{subfigure}[]{0.27\textwidth}
	\centering
	\includegraphics[width=5.3cm,keepaspectratio]{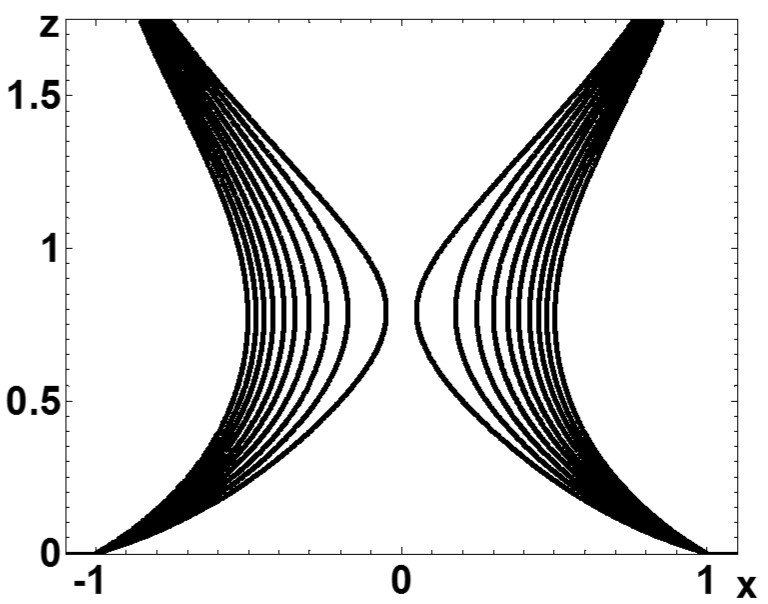}
	\caption{$Oh=2.24\times10^{-3}$ and $\Delta t=17.6.$}
	\label{fig:neck955}
	\end{subfigure}
	\qquad
	\begin{subfigure}[]{0.27\textwidth}
	\centering
	\includegraphics[width=5.3cm,keepaspectratio]{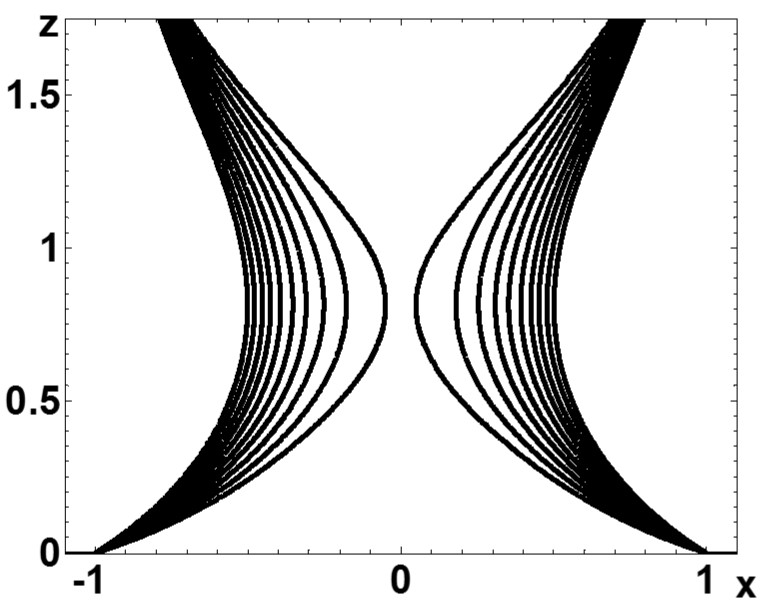}
	\caption{$Oh=10^{-1}$ and $\Delta t=0.67.$}
	\label{fig:neck953}
	\end{subfigure}
	\qquad
	\begin{subfigure}[]{0.27\textwidth}
	\centering
	\includegraphics[width=5.3cm,keepaspectratio]{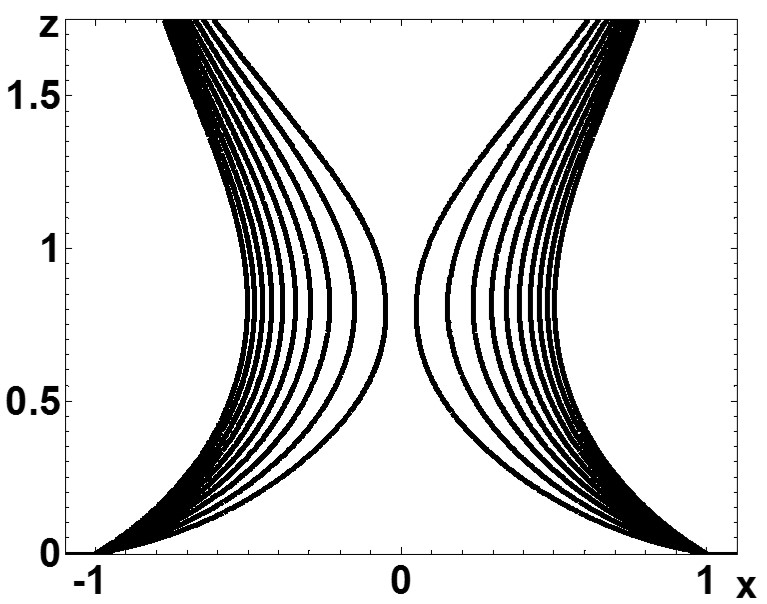}
	\caption{$Oh=10$ and $\Delta t=0.25.$}
	\label{fig:neck951}
	\end{subfigure}
	\caption{The influence of Ohnesorge number on the temporal evolution of the neck of the bubble during the pinch--off process for the case $r_c=1$ and $Q=10^{-3}.$ The outermost and innermost curves represent the respective cases of $r_{min}=r_c/2$ and $r_{min} = r_{tol}.$ The intermediate curves are equally spaced in time, with time step $\Delta t.$ }
	\label{fig:pinch}
\end{center}
\end{figure}
Figure~\ref{fig:pinch} shows the influence of the Ohnesorge number on the pinch--off process. The outermost and innermost curves on each plot show the free surface as $r_{min}=r_c/2$ and $r_{min} = r_{tol},$ respectively. The eight curves in between these are equally spaced in time with time step $\Delta t=17.6,$ $0.67$ and $0.25$ for $Oh=2.24\times10^{-3},$ $10^{-1}$ and $10,$ respectively.

The spacing of the curves show that the pinch--off of the bubble accelerates as $r_{min}\rightarrow 0,$ whilst the radius of curvature at the point on the free surface which represents the minimum neck radius increases with increasing Ohnesorge number. 

\begin{figure}[]
\begin{center}
\includegraphics[width=0.3\textwidth]{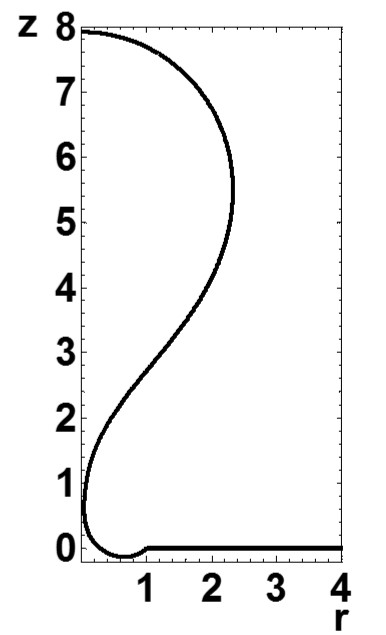}
\caption{Evidence of the liquid phase entering the mouth of the orifice as $r_{min}=r_{tol}$ for the case $(r_c,Oh,Q)=(1,10,7.5)$ where $t_d=6.596$ and $V_d=50.12.$}
\label{fig:weep}
\end{center}
\end{figure} 
Finally, for larger Ohnesorge numbers at sufficiently high flow rates when $r_c=1$, the liquid phase enters the mouth of the orifice towards the end of the formation process (see Figure~\ref{fig:weep}). The gas phase with spatially uniform pressure $p_g$ is present not just above the orifice but below it and the solid surface too. Therefore there is no restriction on the free surface entering the mouth of the orifice if the normal stress boundary condition dictates this. 

\section{Concluding Remarks}\label{sec:disc}
It was shown that the numerical platform based on the finite element method produced benchmark calculations for the formation of a single bubble from a submerged orifice. The global characteristics of the process, the formation time and bubble volume, were found for a wide range of orifice radii, Ohnesorge numbers and volumetric gas flow rates. 

The results were seen to agree very well with experiments, as the finite element method allows the free surface of the bubble to be tracked explicitly and accurately, even when the liquid phase enters the mouth of the orifice, while the computational mesh can be refined in the regions of the flow domain that require extra precision. The simulations could also capture the very small time scales associated with the pinch--off of the bubble whilst the experiments were unable to do so.   

In agreement with the literature, two regimes of bubble formation were identified, the low gas low rate `static' regime and the high gas flow rate `dynamic' regime. The benchmark calculations were used to validate a range of scaling laws proposed in the literature. 

The liquid phase was assumed to perfectly wet the solid surface throughout the entire formation process, and so the three phase solid--liquid--gas contact line remained pinned to the rim of the orifice, whilst the gas was assumed to be inviscid. The model could be extended by relaxing these assumptions in order to investigate the influence of partial wetting conditions, whereby the contact line is free to move along the solid surface, and the viscosity of the gas on the bubble formation process. 

The process of the break up of the bubble, where the topology of the domain occupied by the gas changes, presents a major problem for the modelling. In the present work, it is assumed to have a local effect which is an acceptable approximation for the relatively large bubbles considered here. In a general case, however, especially in microfluidics, it becomes necessary to model the break up process in a singularity--free way. Currently, this problem is outstanding and requires further research. \\

\appendix
\section{Rescaling the Results}
\label{app:rescaling}
In order to convert the results into the dimensionless system used in some works on bubble dynamics, recall that $\bar{r}_c=Lr_c$ and $\bar{Q}=L^2UQ$ are the respective dimensional orifice radius and volumetric gas flow rate. In some works $\bar{r}_c$ and $\bar{Q}/\bar{r}_c^2$ are used as the length and velocity scales rather than those used here of $L=\sqrt{\sigma/\rho g}$ and $U=\sigma/\mu,$ respectively. 

Rescaling in this manner leads to a different group of dimensionless parameters, which will be denoted with hats. Rather than $\left(r_c, Oh, Q\right),$ there is $\left(\hat{Bo},\hat{Oh},\hat{We}\right)$ or $\left(\hat{Bo},\hat{Oh},\hat{Ca}\right),$ where $\hat{Bo}=\rho g \bar{r}_c^2 / \sigma,$ $\hat{Oh}=\mu/\sqrt{\rho \bar{r}_c \sigma},$ $\hat{We}=\rho \bar{Q}^2 / \sigma \bar{r}_c^3$ and $\hat{Ca}=\mu \bar{Q} / \bar{r}_c^2 \sigma,$ are the rescaled Bond, Ohnesorge, Weber and capillary numbers. The rescaled dimensionless formation time and volume are denoted by $\hat{t_d}$ and $\hat{V_d}.$ Then, to convert the parameters used here into the rescaled parameters, we have,\\
\begin{eqnarray*}  \hat{Bo} &=& r_c^2,\\ \hat{Oh} &=& Oh/\sqrt{r_c}, \\ \hat{We} &=& Q^2/Oh^2 r_c^3,\\ \hat{Ca} &=& Q/r_c^2, \\ \hat{t_d} &=& Q t_d/r_c^3, \\ \hat{V_d} &=& V_d/r_c^3, \end{eqnarray*} 
and the inverse is,
\begin{eqnarray*} r_c &=& \sqrt{\hat{Bo}}, \\ Oh &=& \hat{Oh}~\hat{Bo}^{\frac{1}{4}}, \\ Q &=& \hat{Bo}~\hat{Oh}\sqrt{\hat{We}} = \hat{Bo}\hat{Ca}, \\ t_d &=& \sqrt{\hat{Bo}}~\hat{t_d} / \hat{Oh} \sqrt{\hat{We}} =  \sqrt{\hat{Bo}}~\hat{t_d} / \hat{Ca}, \\ V_d &=& \hat{Bo}^{\frac{3}{2}}\hat{V_d}.\end{eqnarray*}

\noindent{\bf{Acknowledgements}}\\





\noindent{\bf{References}}
\bibliographystyle{elsarticle-harv}
\bibliography{<your-bib-database>}

\end{document}